\documentstyle[12pt,aasms4]{article}

\begin{document}

\title{An extinction study of the Taurus Dark Cloud Complex}
\author{H\'ector G. Arce\footnote{National Science Foundation
Minority Graduate Fellow} \& Alyssa A. Goodman\footnote{National Science
Foundation Young Investigator} }
\affil{Harvard--Smithsonian Center for Astrophysics, 60 Garden St.,
Cambridge, MA 02138}
\authoremail{harce@cfa.harvard.edu, agoodman@cfa.harvard.edu}
\vskip0.5in
\centerline {Accepted by {\it The Astrophysical Journal}}

\begin{abstract}
We present a study of the detailed distribution of extinction in a region
of the Taurus dark cloud complex.
Our study uses new $BVR$ images of the region, spectral classification
data for 95 stars, and IRAS Sky Survey Atlas (ISSA) 60 and 100 \micron \/
images.
We study the extinction of the region in four different ways,
and we present the first inter-comparison of all these methods, which are:
1) using the color excess of background stars for which spectral types are
known;
2) using the ISSA 60 and 100 \micron \/ images;
3) using star counts; and 4) using an optical ($V$ and $R$) version
of the average color excess method
used by Lada et al. (1994).
We find that all four methods give generally
similar results--with important exceptions.
As expected, all the methods show
an increase in extinction due to dense dusty regions
(i.e. dark clouds and IRAS cores), and a general increase in
extinction with increasing declination, due to a larger content
of dust in the northern regions of the Taurus dark cloud complex.
Some of the discrepancies between the methods are caused by assuming
a constant dust temperature for each line of sight in the ISSA
extinction maps and not correcting for unexpected changes
in the background stellar population (i.e. the presence of
a cluster or Galactic gradients in the stellar density and average
 \vr \/ color).
To study the structure in the dust distribution, we compare the
ISSA extinction and the extinction measured for individual stars.
From the comparison, {\it we conclude that in the relatively
low extinction regions studied, with $0.9 < A_V < 3.0$ mag
(away from filamentary dark clouds
and IRAS cores), there are no fluctuations
in the dust column density greater than 45\%
(at the 99.7\% confidence level), on scales smaller than 0.2 pc.}
We also report the discovery of a previously unknown open cluster
of stars behind the Taurus dark cloud near R.A 4h19m, Dec. 27\arcdeg30\arcmin
\/ (B1950).
\end{abstract}


\section{Introduction}

In order to understand how molecular clouds evolve and eventually
produce
stars, it is necessary to study the distribution of their star-forming
matter.  Since
the clouds' main constituent, molecular hydrogen, is generally
unobservable, it is
necessary to use other tracers, whose abundance relative to hydrogen can be
reliably
estimated, to map out the distribution of material.
The extinction of background starlight is the result of the  absorption and
scattering of photons off dust grains; so, for a given line of sight, the
amount of
extinction is directly proportional to the  amount of dust. If the
gas-to-dust ratio is known and constant (e.g. Bohlin et al. 1978), then a
detailed study of the dust
distribution in a cloud serves as a detailed study of its mass
distribution.

The study of fluctuations in the dust distribution is also interesting
independent
of its usefulness as a mass tracer.  Strong fluctuations in the
dust distribution have considerable impact on both the physics and
chemistry of the interstellar medium (ISM),
which both depend heavily on the extinction (opacity) structure
on all scales (see Thoraval et al. 1997, and references therein).  In addition,
knowledge of the spatial structure and amount of extinction in the Galactic ISM
is important as it
effects the apparent colors of background sources, such as stars and galaxies.

The most direct measure of reddening is the color
excess of a star with known spectral type.  Unfortunately, mapping out
extended distributions of extinction (reddening) by obtaining photometry and a
spectrum for large numbers of stars is very tedious and time consuming, and
usually impractical.  Therefore, the traditional way of undertaking an
extinction
study of a fairly large region of the sky has been, until recently,
through the use of optical
star counts using photographic plates (Bok \& Cordwell 1973).
This method can only be used
up to an extinction of approximately 4 mag, with a resolution of
$\sim$2.5\arcmin. Thankfully, recent advances in technology have led to the
development of new methods of deriving the  extinction in dark cloud
regions.  For
example, in a study of the structure of nearby dark clouds, Wood et al.
(1994) use  60 and 100 \micron \/ images taken by IRAS to calculate 100
\micron\ optical depth, from which they obtain the extinction ($A_V$).
Lada et al. (1994; hereafter LLCB) took advantage of the improvements in
infrared array cameras to devise a clever new  method
of measuring extinction. The LLCB technique,\footnote{This technique has been
called the ``NICE" ({\bf N}ear {\bf I}nfrared {\bf C}olor {\bf E}xcess)
method by
Alves et al. (1998).} which combines measurements of near-infrared ($H$ and
$K$) color excess and certain techniques of star counting,  has a higher
angular
resolution and can probe greater optical depths than that achieved by optical
star counting alone.

In this study we use four different methods of measuring $A_V$, utilizing:
1) the color excess of individual background stars for which we could obtain
spectral types; 2) ISSA 60 and 100 \micron \/ images to estimate dust
opacity; 3) traditional star counting; and 4) an optical ($V$ and $R$)
version of the average color excess method used by Lada et al. (1994).
To our knowledge, this is the first time that all of these different
methods have
been directly intercompared. We describe the acquisition and reduction of the
data in
Section 2. In Section 3 we present the results of the observations, and in
Section 4
we offer analysis and discussion.  Readers interested primarily in
intercomparison of the various methods and limits on density fluctuations should
skim sections 2 and 3 and focus more on Section 4 and 5. In Section 5 we compare
and rate the four different methods of obtaining
$A_V$. We devote Section 6 to our conclusions.

\section{Data}
The new photometric and spectroscopic observations used in this paper were
originally obtained to conduct the polarization-extinction study described
in Arce et
al. (1998).  The photometry consists of $B$, $V$ and $R$ CCD images of two 10
arcmin by $\sim$ 5 deg  ``cuts'' through the Taurus dark cloud
complex (see Figure 1).  In the spectroscopic observations, we observed 95
stars
(most of them in  cut 1), in order to determine their spectral types.
The cuts  shown in Figure 1 pass through two well known filamentary dark
clouds
(L1506  and B216-217, both at a distance of 140 pc from the Sun)
as well as very low extinction
regions, giving our photometric observations a fairly large dynamic range in
extinction. In the spectroscopic observations, we selected our target stars
along the two cuts by virtue of their relative brightness, in that we
attempted to
exclude foreground stars by not selecting stars which appear unusually bright.
Table 1 lists the spectral type, apparent $V$ magnitude and $B-V$ color,
and the
derived spectroscopic parallax distance for each target star. Our stellar
sample
has only one star with a distance less than 140 pc, which confirms that we
largely
succeeded in excluding foreground stars. In addition to the new photometric and
spectroscopic observations we also obtained co-added images of flux density
from
the IRAS Sky Survey Atlas (ISSA), in order to examine the far-infrared emission
from dust in the region.

\subsection{Photometric Data}
The broad band imaging data of the two cuts (Figure 1)
through the Taurus dark cloud complex were obtained using the  Smithsonian
Astrophysical Observatory (SAO)  AndyCam on the Fred Lawrence Whipple
Observatory (FLWO) 1.2-meter telescope on Mt.~Hopkins,  Arizona.  AndyCam
is a
camera with a thinned back-side illuminated AR coated Loral $2048 \times 2048$
CCD chip. All the frames were taken in $2 \times 2$
bin mode, giving a plate scale of 0.63 arcsec per pixel. In 1995 November, a
total of  64 frames in different positions in the sky
were taken in the $B$, $V$ and $R$ bands, where $R$ is the Cousins $R$ band
filter with an effective wavelength equal to 0.64 \micron.
In each position we obtained one 200 second exposure for each broad-band
filter. Each telescope pointing was a little less than 10\arcmin
\/ north of the previous position, and since each frame is
slightly larger than 10\arcmin \/ $\times$ 10\arcmin, there
is a small sky overlap between the frames successive positions.
The first cut extends from
declination of 22\arcdeg30\arcmin \/
to 28\arcdeg20\arcmin, centered on right ascension 4h 22m 36s (J2000), with
a total of 36 frames.  The second cut extends from declination 24\arcdeg
00\arcmin \/ to 28\arcdeg30\arcmin, centered on right ascension
4h 21m 29s (J2000), with a total of 28 frames. In addition to the
frames acquired in November 1995, seventeen frames
were taken in the $U$, $B$, and $V$ bands in October 1996,
four frames were
taken in the $B$ and $V$ bands in November
1996, and two frames were taken in the
$B$ and $V$ bands in March 1997 ---all with the same instrument configuration
as the November 1995 frames. The additional frames were taken
because: not all of the original frames were of good quality; several
frames were of regions of special interest around the 2 dark clouds
peripheries (outside the cuts)
which the original frames did not include; and because shorter
exposure images of some of the regions covered by the original frames were
needed. The exposure times were 80, 150, and 200 sec for $V$, $B$, and
$U$, respectively, for frames of new sky positions, and 30 sec in $V$, 50
sec in
$B$, and 100 sec in $U$ for frames with repeated positions in the sky.

All of the stellar photometric data reduction was done using standard Image
Reduction and Analysis
Facility (IRAF) routines.  For stars whose spectra had been measured (see
\S 2.2), photometry was obtained in the flat-fielded,  background-subtracted
images using the APPHOT routine.
After analyzing  the dependence of
magnitude value with aperture size in the standard stars  for all nights,
it was decided to use an
aperture radius of 14 pixels. With this aperture size, less than 10\%
of the stars in the most crowded field, with $R$-band apparent magnitude
($m_R$)
between 14.5 and 18.0 mag, have neighbors within the aperture.
The correction to the standard photometric system  was done
using Landolt standards
(Landolt 1992). A set of standards were observed for each night, at
different times of the night, at
varying airmasses. These were then used to solve a set of linear equations
that would give the stars'
$V$ magnitude, \bv \/ and \vr \/  colors using the routines in the IRAF
package PHOTCAL. The errors
obtained from the APPHOT routine, and the errors in the transformation
equation fit were summed in
quadrature  to give the final errors in the photometry. These were $\pm
0.02$ to 0.06 mag for $V$ and
\bv \/ for stars with $V$ between 12.3 and 17.4 mag. We calculated  the
photometry of the standard
stars used to derive the transformation equation  and compared our results
with those quoted by Landolt
(1992). By doing so we  convinced ourselves that the value of our final 1
$\sigma$  errors (Table 1) are a
reasonably good estimation of the true photometry uncertainties.

In addition to obtaining apparent magnitudes and colors for stars, the
photometric data were also used
to do a star count of the region.  The routine DAOFIND was used to detect
sources in the $R$ filter
frames. This routine automatically detects  objects that are above a
certain intensity threshold, and
within a limit of sharpness and roundness, all of which the user specifies. We
used a finding threshold of
5 times the rms sky noise in each frame.  The objects detected by the
routine are then stored in a file
with the objects' coordinates. Although care  was taken to select a limit
of sharpness and roundness so
that DAOFIND would only detect stars, other objects (like cosmic rays and
galaxies) were  also
detected, and some stars clearly above the threshold level were not
detected. Thus, the $R$ frames were
painstakingly inspected visually to erase detected objects that were not
stars, and add the  few stars
that were clearly above the threshold, but were not originally detected.
We obtained the photometry of all the stars in the sample and
then made a histogram (Figure 2)
of the number of stars versus apparent $R$
magnitude in order to study the completeness of the sample. From Figure 2
we estimate the upper completeness limit to be $m_R\approx 18.0$ mag.
Stars brighter than 14.5 mag in the $R$-band 200 second exposures
are saturated, thus the photometry of stars with $m_R < 14.5$ is
unreliable.  Therefore we estimate that our sample is 90\% (or more)
complete
for stars with $14.5 \leq m_R < 18.0$.

\subsection{Spectroscopic Data}
 The spectra of 95 stars along the cuts were obtained using the SAO FAST
spectrograph on the FLWO 1.5-meter telescope on Mt. Hopkins, Arizona.
The observations were carried out during the Fall trimesters of 1995 and
1996. FAST was used with a 3\arcsec \/ slit and a 300 line mm$^{-1}$
grating.  This resulted in a resolution of $\sim$ 6 \AA, a spectral
coverage of $\sim3800$ \AA \/ (from approximately 3600 to around 7400 \AA),
a dispersion of 1.47 \AA/pixel and 1.64 pixels/arcsec along the dispersion
axis.

The spectrum of each star was used to derive its spectral type. In order to
spectroscopically classify the stars, we followed O'Connell (1973) and
Kenyon et al.
(1994) and computed several absorption line indices from the spectra:
\begin{equation}
I_{\lambda} = -2.5 ~ {\rm log} \left[ \frac{F(\lambda_2)}{F'(\lambda_2)}
\right]
\end{equation}
where
\begin{equation}
F'(\lambda_2)= F(\lambda_1) + \left[ F(\lambda_3) - F(\lambda_{1}) \right]
\left[ \frac{\lambda_2 - \lambda_1}{\lambda_3 - \lambda_1} \right]
\end{equation}
is the interpolated continuum flux at the feature,
$\lambda_1$ and $\lambda_3$ are continuum wavelengths, $\lambda_2$  is the
feature wavelength, and $F(\lambda_{i})$ is the average  flux in erg
cm$^{-2}$ s$^{-1}$ \AA$^{-1}$ over a bandwidth specified in Table II of
O'Connell (1973).   We measured the Ca II H ($\lambda 3933$), H$\delta$ \/
($\lambda 4101$), CH ($\lambda 4305$), H$\epsilon$ \/ ($\lambda 4340$),
H$\zeta$ \/ ($\lambda 4861$), and Mg I ($\lambda 5175$) indices of our
program stars and then
compared them to the indices of main sequence stars in the Jacoby et al.
(1984) atlas. The indices have errors of $5-10\%$ depending on the
signal to noise of the spectrum.
This method
resulted in spectral classification  of most of the stars observed with
accuracy of $\sim \pm 1 - 2$
subclasses for spectral types A through F and $\pm 2-4$ subclasses for
stars with spectral type G.
Stars with spectral types earlier than A0 were not found, and stars later
than G9 were not included in
the sample due to the low accuracy in their spectral classification and
reddening corrections. All the
stars were assumed to be  main sequence stars (luminosity type V). Kenyon
et al. (1994) estimate, and
also obtain, that $\sim$ 10\% of their magnitude-limited sample of A--F
stars in the Taurus region are
giants. If we use this result with our magnitude-limited sample it would
mean that only $\sim$7 stars
of our 69 A--F stars are giants. In addition, the intrinsic \bv \/ color of
main sequence A and F stars
differs by less than 0.05 mag from the intrinsic \bv \/ of luminosity type
III and type II stars. Of the
G stars, no more than 5 out of a total of 26 should be giant stars (Mihalas
\& Binney 1981; Table 4-9).
Hence we are not introducing large errors in the extinction
of each star by assuming that all of the stars
we classified are luminosity type V.

Once each star was classified, its reddening was calculated. Intrinsic \bv
\/ values for each spectral
type were obtained from Table A5 in Kenyon \& Hartmann (1995).  The
observed \bv \/  value, from the photometric study, was then
used to obtain the color excess: $E_{B-V}= (\bv) - (\bv)_{\circ}$, where
(\bv) is the observed color
index and (\bv)$_{\circ}$ is the unreddened intrinsic color of the star. An
error in the stellar
classification of $\pm 1-2$ subclasses in A-F stars
transforms into an error of $\pm
0.04 - 0.05$ mag in $E_{B-V}$, while an error in the stellar
classification of $\pm 2-4$ subclasses in G stars transforms into
an error of $\pm 0.05-0.08$ mag in  $E_{B-V}$.
We assumed that $A_{V}=R_{V}E_{B-V}$, with
$R_{V}$ (the ratio of
total-to-selective extinction) equal to 3.1 (Savage \& Mathis 1979; Vrba \&
Rydgren 1985) ---the
validity of this assumption is discussed below.  Using absolute magnitude
values for each spectral type
from Lang (1991), we then obtained distances for each star (see Table 1).

When we calculated the extinction to each star using a constant value of
$R_V =3.1$, we made the
implicit assumption that the ratio of total-to-selective extinction is
constant for different lines of
sight. In fact, $R_V$ varies  along different lines of sight in the Galaxy
and only has a mean of $\sim
3.1$ (Savage \& Mathis 1979).
In contrast with other regions, the Taurus-Auriga molecular cloud complex
seems to have a fairly constant interstellar reddening
law with $R_V \approx 3.1$
through most of the region (Vrba \& Rydgren 1985; Kenyon et al. 1994).
Our $BVR$ photometry is not ample enough to derive $R_V$ for each line of
sight. We would need observations at shorter wavelengths to be able to
independently obtain the value of the ratio of
total-to-selective extinction for each
line of sight we observed. Thus
we decided to use the ISM (and Taurus)
average of $R_V=3.1$, and to caution the reader
that we do not take the errors
caused by assuming a constant $R_V$ into account when we calculate the
errors in $A_V$. In Section
3.1, we show how the ISSA data confirm that $R_V=3.1$ is a good estimate
of the ratio of
total-to-selective extinction for our region.

\subsection{ISSA Data}
The  IRAS Sky Survey Atlas (ISSA) was used to obtain  images of
flux density at 60 and 100
\micron \/ of the Taurus dark cloud complex.  Our region of interest lies
in two
different (but overlapping) ISSA fields. Each of these is a 500- by
500-pixel image, covering a
12.5\arcdeg \/ by 12.5\arcdeg \/ field of sky with a pixel size of
1.5\arcmin. The maps have units of
MJy Sr$^{-1}$, are  made with gnomic projection, have spatial resolution
smoothed to the IRAS beam at
100 \micron\/  (approximately 5\arcmin), and the zodiacal emission has been
removed from them. The
12.5\arcdeg \/ by 12.5\arcdeg \/ images were cropped in order to keep only
the region of the Taurus
dark cloud complex shown in Figure 1. This resulted in a total of four
different images; two (60 and
100 \micron) images of the northern  part and two images of the southern
part of the map.

Although
they have the zodiacal emission removed, ISSA fields are not calibrated so
that the ``zero point''
corresponds to no emission, so another ``background" needs to be
subtracted. This
background subtraction procedure went as follows. First, the minimum value
of each of the four
12.5\arcdeg \/  by 12.5\arcdeg \/ images (see Table 2) was obtained and
subtracted from them. The
resulting images were then used to obtain an optical extinction ($A_{V}$)
map through a process to be
discussed below. At this point, after just a simple subtraction of the
minimum value in each image,
the north and the south A$_V$ maps (see Figure 1) did not agree within the
errors in the
region of overlap. So, the values used for the background
subtraction were
iterated until the best agreement for the overlap region was found,
while keeping the background subtraction constants less than 1 MJy
Sr$^{-1}$ away from the minimum
flux value of the original 12.5\arcdeg \/ by  12.5\arcdeg \/
ISSA fields. Table 2 lists the
values that were ultimately used for this purpose.
These values resulted in a difference of 0.1
mag between the mean in the
distribution of $A_V$ values in the northern image and the mean in the
distribution of $A_V$ values in the southern image.
Tests using other values for background subtraction showed that the
resultant extinction values did not change significantly for small (less
than $\sim$1 MJy Sr$^{-1}$) changes in the background subtraction
constants.  As
discussed below, the offset between ISSA plates winds up being the
limiting error in determining extinction from ISSA data.

The extinction map was computed from the ISSA images using a method very
similar to that described by
Wood, Myers
\& Daugherty  (1994), and references therein.  Note, however, that in their
study, Wood et al. (1994)
used IRAS images which had not gone through  a zodiacal emission removal
process. They devised their own
zodiacal light subtraction, which they state is not very efficient for
regions near the ecliptic, like
Taurus. One of the regions they study was in fact the Taurus dark cloud
region itself. We believe that
our extinction map is of better quality due to the fact that we use ISSA
images which have a more
elaborate zodiacal light subtraction algorithm.

The 60 and 100 \micron \/ dust temperature,
$T_{d}$, at each pixel in an image can be obtained by assuming  that the
dust in a single beam can be
characterized by one temperature ($T_{d}$),  and that the observed ratio of
60 to 100 \micron \/
emission is due to blackbody radiation from  dust grains at $T_{d}$,
modified by a power-law
emissivity. The flux density of emission at a  wavelength $\lambda_{i}$, is
given by
\begin{equation} F_{i} =\left[ \frac{2hc}{\lambda^{3}_{i}}
\frac{1}{e^{hc/\lambda_{i}kT_{d}}-1}
\right] N_{d}\alpha \lambda_{i}^{-\beta} \Omega_{i}
\end{equation} where $N_{d}$ is the column density of dust grains, $\beta$
is the power-law index of
the dust emissivity, $\Omega_{i}$ is the solid angle at $\lambda_{i}$, and
$\alpha$ is a constant of
proportionality.

In order to use equation 3 to calculate the dust
color temperature  ($T_d$) of
each pixel in the image we
have to make some assumptions. The first assumption is that the dust
emission is optically thin. We
believe this is a safe assumption because in our maps there is not a single
line of sight that could be
optically thick ($\tau_{100} > 1$). In fact, the largest $\tau_{100}$
we find in our processed
images is 0.002.  The second assumption we have to make is that
$\Omega_{60} \cong \Omega_{100}$,
which is true for all ISSA images.  With these two assumptions we can then
write the ratio $R$ of the
flux densities at 60 and 100 \micron \/ as:
\begin{equation} R= \frac{F_{60}} {F_{100}} =0.6^{-(3+\beta)} \left[
\frac{e^{144/T_{d}}-1}{e^{240/T_{d}}-1} \right]
\end{equation}
In order to proceed we need to assume a value of $\beta$.
For now we will assume that
$\beta=1$, and we will discuss the implications of this assumption later
on.  We constructed a look-up
table with the value of $R$ calculated for a wide range of $T_{d}$, with
steps in $T_{d}$ of 0.05 K.
For each pixel in the image, the table was searched for the value of
$T_{d}$ that reproduces the
observed 100 to 60 \micron \/ flux ratio.  Using the dust color temperature, we
then calculate the dust
optical depth for each pixel:
\begin{equation}
\tau_{100} =\frac{F_{\lambda}(100 \/ \micron)}
{B_{\lambda}(\lambda,T_{d})}
\end{equation}
where $B_{\lambda}(\lambda,T_{d})$ is the Planck function and
$F_{\lambda}(100 \micron)$ is the observed 100 \micron \/ flux.

We use equation 5 of Wood et al. (1994) to convert from
optical depth to extinction in $V$:
\begin{equation}
A_{V} = 15.078(1-e^{-\tau_{100}/a})
\end{equation}
where $\tau_{100}$ is the optical depth
and $a$ is a constant with a value of $6.413 \times 10^{-4}$.
This equation relies on the work of Jarrett et al. (1989), who present a plot
(their Figure 8) of the relation between 60 \micron\ optical depth
($\tau_{60}$) and
$A_V$ based on star counts.  Assuming optically thin emission, Wood et al.
(1994)
multiply the Jarrett et al. $\tau_{60}$ values by 100/60 to convert to
$\tau_{100}$ and obtain equation 6, above.  Thus, extinction values obtained
using the ISSA images are subject to the uncertainties in the conversion
equation. But, Figure 8 of Jarrett et al. shows that there is a very
tight correlation between  $\tau_{60}$ and $A_V$ for $A_V \leq 5$ mag, implying
very little uncertainty in the conversion of far-infrared optical depth to
visual
extinction. In the Taurus region under study in this paper, all of the
extinction
values measured are less than 5 mag, so we do not consider any errors caused by
uncertainty in the coefficients of equation 6.

 After all this processing was done, we were left with two extinction maps
(a northern and a southern
one) which had some overlap (see Figure 1).   The area of overlap was
averaged and the north and south
extinction maps were combined to produce a final image (Figure 1) extending
from R.A. of 4h 09m 00s to
4h 24m 30s (B1950) and from  Dec. 22\arcdeg 00\arcmin \/ to 28\arcdeg
35\arcmin \/ (B1950).

An interesting point to note is that several ``elliptical holes'' in the
extinction appear in Figure 1. These ``holes'' are unphysical depressions
in the extinction produced by the hot sources seen in the 60 \micron \/
maps. The 60 \micron \/ point sources (mostly embedded young stars) heat
the dust around them and create a region where there is an excess of hot
dust. This limitation, caused by the low spatial resolution of IRAS and
assuming a single $T_{d}$, will then
create an unphysically low extinction, when calculating the $A_V$ in the
region near IRAS point sources, using the method described above.
Unfortunately, cut 2 (see Figure 1)
passes near two of these
unphysically low extinction areas. (In Figures 4 and 5 we mark the
position of the unreal dip in extinction.)
These unphysical holes in $A_V$ are each
associated with two very close IRAS point sources
(one hole is produced by IRAS 0418+2654 and IRAS 0418+2655,
and the other one by IRAS 0418+2650 and IRAS 0419+2650)
inside the dark cloud B216-217 (see end of
\S 3.2).

As mentioned above, in order to calculate the optical depth we assumed that
the dust emissivity is
proportional to a power law ($\tau_d \propto \lambda^{-\beta}$), with index
$\beta=1$. Though studies differ in the values they
find for $\beta$, there is a general agreement that the emissivity index
depends on the grain's  size, composition, and physical structure
(Weintraub et al. 1991). The general consensus in recent years
has been that $\beta$ has a value most likely between 1 and 2,
 that in the general ISM $\beta$ is close to 2, and in denser regions with
bigger grains, $\beta$ is
closer to 1 (Beckwith \& Sargent 1991; Mannings \& Emerson 1994; Pollack et al.
1994). Our
region of interest has both lines
of sight that pass through low and high density medium with different
environments. Thus, there is no
way we can use a ``perfect'' or ``preferred'' value of $\beta$, as it might
be different for different
lines of sight. In our case we had to choose the same value as
Jarrett et al. (1989) and Wood et al. (1994), which is $\beta=1$,
since we use their results to convert from $\tau_{100}$ to extinction.
The errors introduced by assuming a constant $\beta$
are  hard to estimate, since we do not have any way
to measure how much $\beta$ changes in our region of study. We do not include
these errors in the error estimate of $A_V$ using ISSA images (from now on
$A_{V_{ISSA}}$), but it must be
kept in mind that $\beta$ is not necessarily equal to 1 and that its value
may vary for different lines of sight.

In order to estimate the pixel-to-pixel (random) errors in
$A_{V_{ISSA}}$, we examined a circular area of 800 pixels
centered at 4h 19m,  22\arcdeg24\arcmin \/ (B1950) which appears to have a
constant extinction. We compared pixels that are 1.5 IRAS beams apart and
obtained a standard deviation in the extinction
value of this region
to be 0.06 mag. We use this value as an estimate of the pixel-to-pixel
errors in
$A_{V_{ISSA}}$.

Ultimately, though, we need an estimate of the total error in $A_{V_{ISSA}}$,
not just pixel-to-pixel errors.   As explained above, the north and south
extinction maps give slightly different extinction values for matching pixels
in the region where they overlap.  By fitting a gaussian to a histogram of the
difference in $A_{V_{ISSA}}$ value between the north and the south extinction
map, we find a $1\sigma$ width of 0.11 mag. We use this value as an
estimate of the $1\sigma$ error in $A_{V_{ISSA}}$ caused by uncertainty in the
zero level constants and zodiacal subtraction in the ISSA plates. Adding
this ``plate-to-plate" error in quadrature to the the pixel-to-pixel error
(previous paragraph) gives a total error in $A_{V_{ISSA}}$ of 0.12 mag.
We use this value as an estimate of the error in $A_{V_{ISSA}}$, but we
remind the reader that this error does not include any errors caused
by assuming a constant $\beta$.

\section{Results}
The data described above offer the opportunity to measure $A_V$ in four
different ways: 1) using the 85 stars in cut 1
for which we have color excess data
(see Table 1);
2) using 60
and 100 \micron \/ ISSA images as described in Section 2.4;
3) using star-counting techniques on the $R$-band frames; and 4)
using an optical ($V$ and $R$) version
of the average color excess method used by LLCB,
described in more detail in Section 3.3.

Throughout the paper we  assume that the extinction we calculate,
independent of the way it was
obtained, is produced by the dust associated with the Taurus dark cloud
complex at a distance of
140 $\pm 10$ pc (Kenyon et al. 1994).
The region of Taurus we observed lies at  Galactic
coordinate $l^{II} \approx
172\arcdeg$, $b^{II} \approx -17\arcdeg$.
Thus,  our
stars lie towards lines of sight where there is little  or no dust except
for that associated with the
Taurus dark cloud and, we can safely assume that in the area under study,
virtually all the extinction
is produced by the dust associated with the Taurus dark cloud complex.

\subsection{Extinction from the color excess of stars
with measured spectral types}
Using the color excess and position data for the stars in Table 1
that lie on cut 1 and have a distance larger than 150 pc,
we construct the extinction vs. declination plot shown in Figure 3a.
This technique easily detects the rises in
extinction associated with Tau M1 around Dec $23.65\arcdeg$, L1506 around
Dec 25\arcdeg,  B216-217 around
$26.7\arcdeg$, and the area near the IRAS cores Tau B5 and Tau B11
(hereafter Tau B5-B11) around Dec $27.5\arcdeg$. As in the ISSA $A_V$
map (Figure 1), the extinction obtained from
the color excess of stars ($A_{V_{sp}}$) shows an overall rise in extinction
with increasing declination. Also plotted in Figure 3a are the extinction
values obtained from the
ISSA extinction map ($A_{V_{ISSA}}$) for the same coordinates on the sky
where there is a $A_{V_{sp}}$ point.
The value of each $A_{V_{ISSA}}$ point is obtained
from the pixel nearest to the coordinates of each star with measured
$A_{V_{sp}}$
in cut 1. The error bars, set at $\pm 0.12$ mag for each $A_{V_{ISSA}}$ point,
show the total error (including pixel-to-pixel and plate-to-plate
variations, but not including systematic errors) discussed above.

As a check on the assumed value of $R_V$, we let $R_V$ vary in the
calculation of
$A_{V_{sp}}$ and then calculate $\Delta_{tot}=\Sigma^{N}_{i=1}
\vert R_V E_{B-V} - A_{V_{ISSA}} \vert_{i}$ (where $i$ represents the $i$th
point in the plot) for different values of $R_V$. We found that $R_V \sim
3.05$ minimizes the difference between the two curves ($\Delta_{tot}$).
This reassures us that the choice of a constant $R_V =3.1$ is a good
assumption.

The traces of $A_{V}$ vs. declination in Figure 3a show a very
striking similarity. Although the beam size of ISSA is
$\sim 5\arcmin$ and $A_{V_{sp}}$ has
a beam, due to seeing effects,
of approximately 3\arcsec \/ (which transform into 0.2 pc and 0.002 pc,
respectively, at a distance of 140 pc), both values agree within the errors in
most places.
A plot of ($A_{V_{sp}} - A_{V_{ISSA}}$) vs. declination is shown in
Figure 3b.
It can be seen that $A_{V_{sp}}-A_{V_{ISSA}}\approx 0$, within errors, for
all points not in the
vicinity of a steep increase in extinction (i.e. away from dark clouds and
IRAS cores).
In other words, in the low extinction regions $A_{V_{sp}}$
is very similar to $A_{V_{ISSA}}$, despite the great difference in the
resolution of the two methods. This leads us
to believe that there are no or only very small fluctuations in the extinction
inside a 5\arcmin \/ beam, in low $A_V$ regions. In order to put an upper
limit
to the magnitude of the fluctuations, we present a plot of
$A_{V_{sp}} - A_{V_{ISSA}}$ divided by $A_{V_{ISSA}}$, in
Figure 3c. We discuss this plot more in \S 4.1.

The gently sloping solid lines in Figures 3b and c are unweighted linear fits
to the points in each of the plots. Both fits have a very small, but
detectable,
slope  (0.1 mag/degree for Figure 3b).
Previous studies using IRAS data have attributed the existence
of gradients like these to imperfect zodiacal light subtraction
(Wood et al. 1994).
Moreover, the fact that the linear fit of the middle panel crosses the
$A_{V_{sp}} - A_{V_{ISSA}}$ zero line at a declination of around 25.4\arcdeg,
near the middle of the overlap region between the northern and southern
$A_{V_{ISSA}}$ (Figure 1)
map, leads us to believe that the small gradient is due
to imperfect ISSA image reduction.
The gradient in $A_{V_{ISSA}}$ gives a pixel-to-pixel offset of 0.008 mag/beam,
which is much less than the pixel-to-pixel random errors in $A_{V_{ISSA}}$, and
much, much less than other systematic errors (e.g. constant $\beta$) so we
do not
to correct for it.

\subsection{Star counting}
With the help of IRAF, as described in \S 2.1, we located a total of
3,715 stars
in cut 1 and 3,074 stars in cut 2, with $14.5 \leq m_R < 18.0$,
from the November 1995 $R$-band images.
With this database of stellar positions we measure the extinction
over the region covered by both cuts, using classical star counting
techniques.
First, we
subdivide each cut into a rectilinear grid of
overlapping squares,
and then we count the total number of stars in each square.
Our ultimate goal in star counting is to obtain the extinction of the region
in a way that can be compared to the other techniques used in this
paper. Therefore, in order
to mimic the resolution and sampling frequency of the ISSA data, we
made the counting squares 5\arcmin \/ on a side (the resolution of ISSA)
and the centers
of the squares were separated by 1.5\arcmin \/ (the size of an ISSA pixel).

Conventionally,  measuring extinction from star counts involves comparing
the integrated number of stars within a given cell
towards the region of interest to a nearby reference field which is assumed
to be free from extinction (Bok
\& Cordwell 1973). Several assumptions need to be made in order to use this
method:
1) the population of stars background to the region of interest does not vary
substantially and is
similar to the reference field; 2) the extinction ($A$) is uniform over the
count
cell; and 3) the integrated surface density of stars for the
reference field ($N_{ref}$)
of stars brighter than the apparent magnitude, $m$,
follows an exponential law with
${\rm log}(N_{ref}(<m))=a+bm$.
The integrated surface density of stars for any
other field under study, $N_{on}$,
follows a similar law, ${\rm log}(N_{on}(<m))=a+b(m-A)$.

In our case, the extinction to each square was obtained
via:
\begin{equation}
A_{V_{on}} = A_{V_{ref}} + \frac{A_{V}}{A_{R}}
\frac{{\rm log}(n_{ref}/n_{on})}{b_{R}}
\end{equation}
where $A_{V_{ref}}$ is the extinction of the reference field,
$n_{ref}$ is the number of stars in the reference field,  and $n_{on}$
is the number of stars in any of the other counting cells. The quantity
$b_{R}$ is the slope of the cumulative number of stars as a function of
$R$ apparent magnitude (see assumption 3 in previous
paragraph). We calculate $b_R$ by fitting a line to
${\rm log}(N_{ref}(<m_R))$ for $14.5 \leq m_R<18.0$, and obtain a value of
$0.34 \pm 0.02$.
The ratio
$A_{V}/A_{R}$ is the reddening law between $V$ and $R$
wavelengths, which gives the value of 1.24  (He et al. 1995)
used in the conversion for star counts in (Cousins) $R$ to extinction in $V$.

As mentioned above, in conventional star count studies, the reference fields
are areas in the sky, close to the region under study, which
have $A_{V_{ref}} = 0$.
This is not so in our case.
Since our CCD data were not originally obtained for star counting purposes,
we did not take any images of a
reference field with $A_{V_{ref}} = 0$. Thus, the reference fields were
chosen to be regions where we could safely assume the value of $A_{V_{ref}}$.
Specifically, the reference fields were chosen by virtue of their having a
spectroscopic probe star to which we derived
the extinction through its color excess.

One of the major assumptions in conventional star counting
studies is that the population of background stars to the cloud is similar
to that in the reference field. This is why star count studies are only
done over small regions of the sky. In our case,
cut 1 (the longer cut) expands 6 degrees in declination which transforms to
nearly 4\arcdeg \/ in Galactic latitude ($-18.8\arcdeg <b^{II}<
-14.9\arcdeg$).
This span in Galactic latitude is enough to have a big effect on the
surface density of stars due to galactic variations;
one detects more stars per cell, for a constant $A_V$, the closer one
observes to the galactic plane. (The star count Galaxy model of Reid et al.
[1996]
predicts that a 10 square degree field with no extinction
centered at $b^{II}=-14\arcdeg$,
$l^{II}=-172\arcdeg$ will have approximately a factor of two more stars than
a  10 square degree field with no extinction
centered at $b^{II}=-19\arcdeg$, $l^{II}=-172\arcdeg$.)
Thus, it was imperative that we correct
for Galactic changes in the stellar density. Not doing this would have resulted
in an unreal drop in the derived extinction for lower Galactic latitudes.
We account for the Galactic gradient by using 4 different, more or less
evenly spaced, reference fields at
different galactic latitudes along the cut.
Each of these fields has a star with measured reddening (from
Table 1) which was used to estimate the extinction of the reference field in
question. Thus, for example, the star count extinction
located between $b^{II}=-18.9\arcdeg$ and $b^{II}=-17.9\arcdeg$
(which transforms to $22\arcdeg25\arcmin<\delta_{J2000}<23\arcdeg56\arcmin$
 in cut 1)
is tied to the reference field where star 011001 lies; the
star count extinction in the region with $-17.9\arcdeg\leq b^{II}<-16.9\arcdeg$
(which transforms to
$23\arcdeg56\arcmin<\delta_{J2000}<25\arcdeg25\arcmin$ in cut 1)
is tied to the reference field where star 051043 lies; and so forth.
The number of stars ($n_{ref}$), the visual extinction ($A_{V_{ref}}$),
and the range in $b^{II}$ which each of the four reference
fields calibrates are given in Table 3.
It is important to stress that by doing this calibration we are tying
$A_{V_{sc}}$ to $A_{V_{sp}}$ at these four points, and thus $A_{V_{sc}}$ is
not totally independent of $A_{V_{sp}}$. But this procedure only
forces $A_{V_{sc}}$ to agree in absolute value to $A_{V_{sp}}$ in four points
and not to give the same structure or scale through the cuts.

In our study, where observations lie primarily along a declination cut,
the best way to graphically  compare the extinction measured
by star counting ($A_{V_{sc}}$) and from the ISSA images
($A_{V_{ISSA}}$) is to plot them both in an extinction versus
declination plot. (Cut 1 is about 6 degrees long, but only 10\arcmin
\/ wide, giving a ratio of 1:36 between length and width. This makes
it practically impossible to show a  legible figure of a
star count extinction map of the cuts.)
A value of the extinction was obtained for each
$5\arcmin \times 5\arcmin$ counting box, and
then averaged extinction over R.A. for every point in declination,
so as to produce only one value of $A_V$ for each declination,
independent of R.A.
Constant declination slices  every 5\arcmin \/ show
that the variations in  $A_{V_{ISSA}}$ across the
10\arcmin \/ width of each cut  are not large. Most of the
constant declination slices
had standard deviations in $A_{V_{ISSA}}$ of less than 0.2 mag,
and none exceeded 0.3 mag. Moreover, most adjacent counting
boxes, with the same declination, differ in $A_{V_{sc}}$ by less
than 0.2 mag.
Therefore, we are confident that we are not introducing large errors by
averaging the extinction over the $\sim$ 10\arcmin \/  spanned by each
cut in Right Ascension. On the other hand, by averaging
over R.A., the sensitivity to small scale fluctuations decreases.
The smoothed (averaged) extinction trace has less sensitivity in detecting
fluctuations on 5\arcmin \/ scales than on
10\arcmin\ scales, where it reaches full sensitivity.
Nevertheless, averaging the extinction over R.A.
does not create any disadvantage for the purpose of comparing
the different ways of calculating the extinction since all the
techniques are averaged over the same width.

Figure 4 shows $A_{V_{sc}}$ and $A_{V_{ISSA}}$ (now both averaged over the
$\sim$ 10\arcmin \/ that spanned each cut in R.A.)
versus declination, for both cuts.\footnote{Notice that $A_{V_{ISSA}}$ in
Figure 4 is an average over
the approximately 10\arcmin \/
width of the frames taken in the optical, but still has resolution
$\sim 5$\arcmin
\/ along the declination direction. Thus, these are not exactly the
same values of $A_{V_{ISSA}}$ shown in Figure 3, where $A_{V_{ISSA}}$
values for
1.5\arcmin \/ single ISSA pixels, with 5\arcmin \/ resolution, at individual
stellar positions are plotted.} The random errors in the star count trace  are
plotted for each point. The uncertainty in the
number of stars in each  sampling box is given by
$\sqrt{n}$ (Poisson statistics), where $n$ is the  number of stars counted
in the box (Bok 1937).
The uncertainty in the extinction (with contributions from the uncertainty in
$n$, $n_{ref}$, $A_{V_{ref}}$, and $b_R$)
for each sampling box with the same declination is then averaged to give the
final (plotted) error in the extinction. Like the ratio of total-to-selective
extinction ($R_V$), the value of $A_V/A_R$ may vary from one line of sight to
the other. Also, like $R_V$ in section 2.2, here we use the ISM average
value of $A_V/A_R$. We do not take the errors caused by assuming
a constant value  of $A_V/A_R$ into account when we calculate the errors
in $A_{V_{sc}}$, as we have no way of calculating them.

It can be seen that both $A_{V_{ISSA}}$ and $A_{V_{sc}}$ show the same gross
extinction structure. Both have local maxima and minima in the same places (in
most cases), but not at necessarily the same value. Both traces detect rises in
extinction associated with Tau M1, L1506, B216-217, and Tau B5-B11. It is clear
that
$A_{V_{sc}}$ has more fluctuations than the $A_{V_{ISSA}}$ trace. These
fluctuations are not likely to be real, as most of them are of the same
magnitude
as the errors in $A_{V_{sc}}$. Most of the ``noise'' in the  extinction is
due to
the fact that the  star count technique is very dependent on the assumption
of a
constant background stellar surface density. Real (small) changes in the
background surface density of stars (not caused by extinction) will produce
unreal fluctuations in the resultant $A_{V_{sc}}$.

Note the unreal dip in the cut 2 trace of
$A_{V_{ISSA}}$ (see Figure 4),
caused by assuming a constant dust temperature for those lines of sight
where there are IRAS
point sources which heat the dust around them (see section 2.3).
This shows the potentially large errors in $A_{V_{ISSA}}$ that can arise
from assuming a single dust temperature ($T_d$) for each line of sight.
The other place where there is a significant discrepancy between
$A_{V_{ISSA}}$ and $A_{V_{sc}}$ is in the rise in extinction
associated with Tau B5-B11 (see Figure 4), where the two methods
disagree by more than $1\sigma$ of the error in $A_{V_{sc}}$.
This discrepancy will be discussed further in \S 4.2.

\subsection{Extinction measured via average color excess method}
Taking advantage of the fact that we had taken our November 1995 images in
more than one broad band filter,
we used the method developed by LLCB to study the extinction along
both cuts in yet another way.  This method consists of assuming that the
color distribution of stars
observed  all along the cut is identical in nature to that of stars in a
control field.
With this assumption one can use the mean $V-R$ color of the stars in the
control field to approximate the intrinsic $V-R$ color of all stars
background to the cloud. (Note that LLCB's analysis was in the
near-infrared, and they used $H-K$
colors---which have even less intrinsic variation than $V-R$ colors.) Using
the same technique as in
star counting, the region under study is divided into a  grid of
overlapping counting boxes, and then an
average of the color excess (and extinction) of the stars in each counting
box is obtained.

To derive extinction from the average color excess method, we used the same
sample of stars, with $14.5 \leq m_R < 18.0$, used in our star counting
study (\S
3.2). The region between declinations 22.9\arcdeg \/ and 23.22\arcdeg \/
(B1950) was used as our
reference field since, as can be seen in Figure 3a, this region has a uniform
extinction within the errors. We took an average of the extinction of the 6
stars for which we had spectral types and which lie inside this region,
and obtained a value of $A_{V_{ref}}=0.72 \pm 0.2$.
We did not divide the area under study into square sampling boxes as it is
usually done in star count studies (see previous section).
Instead, the area under study was divided into rectangular cells,
where the R.A. side of the rectangle was
dictated by the frames' width ($\sim 10\arcmin$) and the declination side
was set to be 5\arcmin.
The centers of the rectangles were separated in declination by 1.5\arcmin.
This was done in order to keep the same resolution and sampling
frequency in the declination direction as in the
ISSA and star counting methods (described in the previous sections).
We are not sensitive to any variations in extinction within each
measurement rectangle, but Figure 3a suggests that such variations are very
small.

The number of stars in each rectangle was counted and the color excess
of each star was obtained using the formula:
\begin{equation}
E_{V-R} = (V-R) - \langle V-R \rangle_{ref}
\end{equation}
where $\langle V-R \rangle_{ref}$ is the mean \vr \/
color of the reference field,
which in our case is $0.60\pm0.05$ mag. Figure 5
shows the
distribution of \vr \/ colors in the reference field.\footnote{Notice that the
width of the color distribution in Figure 5 is significantly larger than the
error in the mean (0.05 mag).  The spread in near-IR (e.g. $H-K$) color for a
field like this would be much narrower, which is why it is preferred to use
the average color excess method in the near-IR (LLCB).} We then apply
equation 8
to each star in a counting rectangle and obtain a mean color excess to each
rectangle:
\begin{equation}
\langle E_{V-R} \rangle = {\frac{\Sigma_{i}^{N} (E_{V-R})_i}{N}}
\end{equation}
where $N$ is the number of stars in a counting rectangle and $(E_{V-R})_i$
is the color excess of the $i$th star.
The mean visual extinction is obtained using:
\begin{equation}
\langle A_V \rangle = A_{V_{ref}} + {\frac{A_V}{E_{V-R}}} \times \langle
E_{V-R} \rangle
\end{equation}
where $A_{V_{ref}}$ is the extinction of the reference field (0.72 mag), and
we use the ISM average value of 5.08 for the expression
$A_V/E_{V-R}$ (He et al. 1995). Similar to the extinction from star counts,
the extinction using the average color excess (from now on $A_{V_{ce}}$)
is not totally independent of  the spectral classification method.
Recall that $A_{V_{ref}}$, which is used to calibrate $A_{V_{ce}}$,is
determined
using measurements of $A_{V_{sp}}$. This calibration forces $A_{V_{ce}}$ to
agree in absolute value with the average of $A_{V_{sp}}$ in the reference
field,
but it does not force $A_{V_{ce}}$ to give the same structure or scale in the
extinction curves throughout the cuts.

Using the average color excess technique,
we are able to detect the same overall trends in extinction
found from the ISSA plates, spectral analysis, and star counting
 (see Figure 6). The rises in extinction due to Tau M1, L1506, B216-217,
and Tau B5-B11 can be seen as well pronounced peaks in $A_V$. Again,
note the unreal dip in the cut 2 trace of $A_{V_{ISSA}}$,
caused by the presence of four IRAS
point sources in the dark cloud B216-217 (see section 2.3).
In Figure 6 we also show the random
errors of each point in the $A_{V_{ce}}$ trace.
The measurement uncertainty in $A_{V_{ce}}$ for any given counting cell is
given by:
\begin{equation}
\sigma_{A_{V_{ce}}}= \sqrt{\sigma_{ref}^2 + (5.08)^2[ \sigma^{2}_{mean} +
\Sigma_{1}^{N} \frac{\sigma^{2}_{(V-R)_{i}}}{N^2}]}
\end{equation}
where $\sigma_{ref}$ is the uncertainty in $A_{V_{ref}}$
(which is equal to 0.2 mag), $N$ is the number of
stars in the counting cell and $\sigma_{(V-R)_{i}}$ is the photometric
error in $V-R$ of the $i$th star inside the cell. The quantity
$\sigma_{mean}$ is the error in the mean of the
$V-R$ color distribution of the counting cell.
The distribution of the $V-R$ colors does not have a gaussian distribution,
thus $\sigma_{mean}$
was obtained using Monte Carlo simulations. We obtained $N_{mc}$ values
of $V-R$, representing the $V-R$ colors of $N_{mc}$ stars in a counting cell,
drawn from a distribution given by the reference field distribution (Figure 5).
We then computed the average $V-R$ color over these 
 $N_{mc}$ stars. The procedure was repeated a thousand times 
with the same $N_{mc}$, from which we obtained a
(gaussian) distribution of average values. The 1$\sigma$ width of this gaussian
was then used as the value of $\sigma_{mean}$.
This procedure was repeated for different values of $N_{mc}$ (representing
different number of stars inside a counting cell).
We do not include the errors in $A_{V_{ce}}$ caused by
assuming a constant $A_V/E_{V-R}$ for all lines of sight.

Although $A_{V_{ce}}$ and $A_{V_{ISSA}}$ agree very well for the low
declination ($\delta < 25.5\arcdeg$) part of both cuts, their values seem to
diverge for
higher declinations ($\delta > 25.5\arcdeg$; Figure 6). This effect suggests that it
may be inappropriate to assume a single average color for the whole length (6\arcdeg) of our cuts.  A change in the average \vr \/ could be due to:
 1) a gradient in
the  $\langle V-R \rangle$ caused by a greater fraction of early type stars
close
to the galactic plane; and/or 2) a sudden drop in the value of $\langle V-R
\rangle$ in the north edge of the cuts due to the presence of a star cluster.
Concerning the first point,
the star count Galaxy model of Reid et al.
(1996) predicts that a 10 degree square degree field with no extinction
centered at $b^{II}=-14\arcdeg$,
$l^{II}=172\arcdeg$ will have approximately an average star ($V-R$) color
0.04 mag greater than
a  10 degree square degree field with no extinction
centered at $b^{II}=-19\arcdeg$, $l^{II}=172\arcdeg$. A difference of 0.04 mag
in $\langle V-R \rangle$ transforms to a difference of 0.2 mag in
$A_{V_{ce}}$.
Thus, it is possible that at least some
of the discrepancy between
$A_{V_{ce}}$ and $A_{V_{ISSA}}$ for the northern parts of both cuts is due
to this uncorrected effect of varying $\langle V-R \rangle$. We will discuss
the possibility of a cluster in section 4.3.

\section{Analysis and Discussion}
\subsection{Structure in the cloud}
Figure 3a shows a striking resemblance between the extinction
obtained through the use of the ISSA 60 and 100 \micron \/
images ($A_{V_{ISSA}}$) and the extinction obtained
through the color excess of individual stars which we had classified by
spectral type ($A_{V_{sp}}$).
Our stellar reddening (color excess) sample
represents a map of the distribution of extinction
($A_{V_{sp}}$) which, although it has a ``pencil beam resolution,''
is measured in a spatially nonuniform fashion.
On the other hand the extinction data obtained from the ISSA images  is
spatially continuous, with
1.5\arcmin \/ pixels and a resolution of  5\arcmin \/ ($\sim 0.2$ pc at a
distance of 140 pc). Most stars with measured $A_{V_{sp}}$ are less than
5\arcmin \/ from their
nearest star with measured $A_{V_{sp}}$,
and there are a number of cases where three and even four
stars lie within 5\arcmin \/ of each other. Therefore if one were to place
a 5\arcmin \/ beam anywhere along cut 1, one would find that 1 to 4 stars
would lie, in random places, inside the beam.
Thus, if there were to exist big fluctuations in the
dust distribution inside the
0.2 pc beam of  IRAS, we would see strong variations between
the value of $A_{V_{sp}}$
and the value of $A_{V_{ISSA}}$. Note that
this fluctuation probe can only be used where there are stars that have been
classified by spectra.

Using the extinction measurements shown in Figure 3, we can place an upper limit
on the fluctuations in the dust distribution within a 5\arcmin\ beam. In the
bottom panel of Figure 3c we plot the difference between
$A_{V_{sp}}$ and $A_{V_{ISSA}}$ divided by $A_{V_{ISSA}}$ (from now on $\Delta
A_V/A_V$) versus declination. Here the errors are obtained using the quoted
errors for $A_{V_{sp}}$ (see Table 1) and
$A_{V_{ISSA}}$ (0.12 mag), and propagation of errors.
One can think of $\Delta A_V/A_V$ as a
measurement of the deviations in the average extinction within
a fixed {\it area} of the sky. In our case the area is given by the 5\arcmin \/
beam of $A_{V_{ISSA}}$.

Figure 3c has only four points with an absolute
value which is more than 3 times its 1 $\sigma_{i}$ error
(independent of whether we correct for the small
gradient in $A_V$ with declination), where $\sigma_{i}$ is the error of
each individual point on Figure 3c.
Each of these four points
is associated with one of the extinction peaks created by dark clouds and IRAS
cores intersecting the cut. The  high values of $\Delta A_V/A_V$ could be due to
two effects indistinguishable by our data; spatially unresolved steep
gradients in the extinction or random
fluctuations in the dust distribution inside dark clouds and
IRAS cores.  These possibilities will each be discussed later.
All of the remaining lines of sight, in the more uniform extinction areas,
have absolute values of  $\Delta A_V/A_V$
which are less than 3 times their 1 $\sigma_{i}$ error.
If we exclude points near dramatic
extinction peaks (see Figure 3), then we do not detect any
deviations from zero in $\Delta A_V/A_V$ within our sensitivity.

We can characterize our sensitivity to extinction fluctuations
using the average error in $\Delta A_V/A_V$,
which is given by $\sigma_{av}=\Sigma_{i}^{N} \sigma_{i}/N$.
The 1-$\sigma_{av}$ error in $\Delta A_V/A_V$,
for points with $A_{V_{ISSA}} < 0.9$ mag is 0.41, while for points
with $A_{V_{ISSA}} \geq 0.9$ is 0.15.
We choose $A_{V_{ISSA}}=0.9$ as the dividing line since points with
$A_{V_{ISSA}} < 0.9$ mag have consistently large uncertainties.
Assuming that 0.15 is the
``typical'' error in $\Delta A_V/A_V$
for points with $A_{V_{ISSA}} \geq 0.9$,
we can then state that
a real detection (using a 3-$\sigma_{av}$ detection limit)
of sub-IRAS beam structure
would be if $\Delta A_V/A_V \gtrsim 0.45$.
Therefore only in places where $\Delta A_V/A_V > 0.45$,
can we say that there are
sub-IRAS beam structures in the cloud.
Any value less than 0.45 would be considered part of the ``noise''
and not significant enough to be considered a detection of sub 0.2 pc
structure. So, ultimately, we only {\it detect} deviations from the mean $A_V$
within a 0.2 pc beam in the vicinity of IRAS cores and dark clouds.  Outside of
those regions, deviations within a 0.2 pc beam are limited to be less than
$\Delta A_V/A_V < 0.45$ (for $A_V>0.9$). For points with extinction less than 0.9
mag the larger uncertainty in the extinction determinations means that only
points with
$\Delta A_V/A_V \gtrsim 1.23$ would be real fluctuations detections, and no
such points are found.

\subsection{Evidence for smooth clouds}

In a very important study, Lada et al. (1998; hereafter LAL) recently showed
that smoothly varying density gradients
can produce the ``fluctuations" observed in extinction studies of filamentary
clouds. Two studies of dust extinction in filamentary dark clouds
had been conducted previous to LAL: 1) the study of IC 5146
by LLCB; and 2) a study of L977 by Alves et al. (1998).
Both studies find that in $1.5\arcmin \times 1.5\arcmin$ cells,
the dispersion of extinction measurements
within a square map pixel (what LLCB name $\sigma_{disp}$)
increases in a systematic way with the average $A_V$,
in the range of $0 < A_V < 25$ mag. Both studies
 conclude that the systematic trend
in their $\sigma_{disp} -A_V$ plot, an increase of $\sigma_{disp}$ with $A_V$,
is due to variations in the cloud structure on scales smaller than the
resolution of their measurements. But neither of the studies could definitively
determine the nature of the fluctuations in the extinction.
This prompted LAL to study IC 5146 in the same way as LLCB, but at a higher
spatial resolution (30\arcsec).
With the help of Monte Carlo simulations
LAL conclude that the form and slope of the $\sigma_{disp} -A_V$
relation, and hence most (if not all) of the small scale variations
in the extinction are due to unresolved gradients in the dust distribution
within the filamentary clouds.  LAL note that random spatial
fluctuations in the dust distribution
could exist, at a very low level, in addition to the smooth gradients.
They state  that $\sigma_{ran}/A_V$ due to random fluctuations is much less
than
25\% at $A_V \sim 30$ mag, which is consistent with our
($3\sigma$) upper limit of $\Delta A_V/A_V =0.45$ at $0.9<A_V<3.0$ mag.

Recently, Thoraval et al. (1997), hereafter TBD,
observed a low $A_V$ area of
the IC 5146 dark cloud complex. Similar to our study,
TBD concentrate their observations in a low and uniform extinction
region ($A_V < 5$), but unlike our study, the region that 
TBD studied did not include  a filamentary
cloud. They conclude that the variations
in the extinction are present at a level no larger than
$\Delta A_V/A_V \sim 0.25$, again consistent with our
($3\sigma$) upper limit of $\Delta A_V/A_V =0.45$, and similar to
what LAL obtain in the high extinction region of IC 5146.
If we exclude, in our study, the points near extinction peaks, we are
left with the same results as TBD: no fluctuations on scales smaller than
the resolution. These results all suggest that there is
very little random spatial
fluctuation in the extinction in regions of low $A_V$ far from
extinction peaks.

Although it is not possible to definitively determine the origin of the
handful of high $\Delta A_V/A_V$ we find near extinction peaks, it
is more likely that they are due to
unresolved steep gradients in clouds
than to localized random fluctuations in the dust distribution.
The filamentary dark clouds and IRAS cores typically
have a minor axis that is only 3 to 4 times the IRAS beam size, so some IRAS beam
will undoubtedly contain a steep extinction gradient characterizing the
the ``edge" of one of these structures. Extinction measurements using a
pencil-beam (e.g. $A_{V_{sp}}$), {\it would} be able to resolve this ``edge,"
so, near edges, large-beam (e.g. $A_{V_{ISSA}}$) and pencil-beam measurements
would disagree. The results of TBD and LAL reinforce this
hypothesis. Thus, we strongly believe that the high values of
$\Delta A_V/A_V$ near dark clouds and IRAS cores are due to
steep gradients in the extinction  not resolved by the IRAS beam.

\subsection{Possible discovery of a previously unknown cluster}
While comparing our different ways of obtaining the extinction we found a
peculiar discrepancy between
$A_{V_{ISSA}}$ and $A_{V_{sc}}$ and between $A_{V_{ISSA}}$ and $A_{V_{ce}}$,
in the declination range from 27.2\arcdeg \/ to 28\arcdeg \/ (B1950)
along the two cuts (see Figure 4). The rise in ISSA
extinction in these declinations is due to the existence of a high dust
concentration which
Wood et al. (1994) classify as the IRAS
cores Tau B5 and Tau B11 (Tau B5-B11).
The trace of $A_{V_{ISSA}}$  shows that the increase
in the extinction associated with Tau B5-B11 is of similar or
higher magnitude to the increase in extinction
associated with the dark cloud B216-217, in both cuts (see Figure 6).
On the other hand, the star count extinction and the average color excess
extinction show a small increase in $A_V$ associated with Tau B5-B11
compared to that associated with B216-217.
In addition, there is no sharp
decrease in the surface density of stars like the one associated with the
two dark clouds in our cuts.
This can be observed in both the spatial distribution of our
$R$-frame stars and in the Digitized Palomar Sky Survey.

One possible explanation for the discrepancy between extinction traces
is that either $A_{V_{ISSA}}$ or $A_{V_{ce}}$ and
$A_{V_{sc}}$ were calculated using the wrong assumptions.
It could be that the dust in the Tau B5-B11 region has different physical
properties compared to the rest of the dust in the Taurus dark cloud
complex,
which  would change the values of $\beta$ or of $A_V/A_R$ and $A_V/E_{V-R}$.
When we calculated $A_{V_{ISSA}}$, $A_{V_{sc}}$, and  $A_{V_{ce}}$,
we assumed that the power law index, $\beta$ (equation 4), $A_V/A_R$
(equation 7)
and $A_V/E_{V-R}$ (equation 10) were constant for all lines of sight.
We could change the value of $A_V/A_R$ (equation 7)
from 1.24 to 1.74 in order for $A_{V_{sc}}$ to be approximately
equal to $A_{V_{ISSA}}$ for the lines of sight that pass through Tau B5-B11.
But, the value of $A_V/A_R$ is tied to the value of $A_V/E_{V-R}$,
by the equation
${\frac{A_V}{E_{V-R}}} = {\frac{1}{1-{\frac{A_R}{A_V}}}}$, thus
the proposed change in $A_V/A_R$ would change $A_V/E_{V-R}$ from 5.08 to 2.35,
making the discrepancy
between $A_{V_{sc}}$ and $A_{V_{ISSA}}$ (Figure 6) more pronounced.
An alternate possibility is that the value of $\beta$ changes from 1 to a
value
less than 1
for lines of sight in the region of Tau B5-B11, in that case
$A_{V_{ISSA}} \approx A_{V_{sc}} \approx A_{V_{ce}}$.
Although it is possible to have neighboring
lines of sight with different dust properties, it is very unlikely
(but not impossible) to have dust with $\beta < 1$
in a region like Tau B5-B11, according to
experimental and theoretical studies of dust properties (Weintraub et al. 1991;
Pollack et al. 1994).

A more likely explanation for the discrepancy
between extinction traces near Tau B5-B11
is that there is a sharp increase in the stellar distribution
in the area, which we did not account for when we calculated
$A_{V_{sc}}$. The existence of a previously unknown stellar cluster in the
vicinity of R.A. 4h 19m, Dec. $27\arcdeg30\arcmin$ \/ (B1950) could create
such an increase in the number of stars in the region.
Also, if the cluster is relatively young, the average stellar colors
should be bluer than in the field.
This can explain why the $A_{V_{ce}}$ trace follows, within the error, the
structure present in the $A_{V_{ISSA}}$ trace, but at a lower value, while
the $A_{V_{sc}}$ trace decreases in extinction value without following
the structure in $A_{V_{ISSA}}$.

We conclude that there is a sudden increase in the stellar distribution
background to
Tau B5-B11 (and a change in the
average \vr \/ color),
which is due to a previously unknown open star cluster.\footnote{The Lynga catalog of star clusters (Lynga 1985)
does not contain a cluster in the vicinity of R.A. 4h 19m, Dec.
27\arcdeg30\arcmin \/ (B1950).}
This is more credible than a change in dust properties, since the cluster
hypothesis does not require the assumption of a physically contradictory,
simultaneous change in the value of $A_V/A_R$ and $A_V/E_{V-R}$,
or an improbable value of $\beta$.
We expect that further observations of the area will verify the existence
of an open cluster behind Tau B5-B11.

\section{Rating the Various Methods}

Although we find that all four methods give generally similar results
and are consistent with each other, there are some important exceptions.
The discrepancies arise from different systematic errors inherent to
the different techniques. For example, when calculating $A_{V_{ISSA}}$,
a constant dust temperature for each line of sight was assumed. This
single temperature assumption breaks down in the immediate vicinity
of stars surrounded by dust. Even though
it is very likely that there is dust of many
different temperatures along the lines of sight to these stars, it is
the hot dust that dominates the emission at 60 and 100 \micron,
resulting in an incorrect estimate of
$\tau_{100}$ (and $A_{V_{ISSA}}$),
as we observe in the regions near IRAS sources.

The techniques that use background stellar populations to
measure the extinction (i.e, star count and the average excess color
method) also suffer from important systematic errors. Here the
major systematic error lies in assuming a constant stellar population
background to the cloud.  If the region under study spans
several degrees in galactic latitude (which is our case) uncorrected
gradients in the background stellar density and the average
stellar color will lead to incorrect extinction measurements when using
for the star count method and the average excess color method, respectively.
In addition, small fluctuations in the background surface stellar density can
result in unreal fluctuations in the extinction.

It is important to appreciate that the techniques used in this Paper are not
entirely standalone or independent methods of obtaining extinction.
Even the most exact method for measuring extinction, using the
color excess of individual stars with measured spectral type ($A_{V_{sp}}$)
still depends on the value of $R_V$.
Both star counting ($A_{V_{sc}}$) and the average color excess ($A_{V_{ce}}$)
technique depend on a reference field of
known extinction for calibration.  In this study, $A_{V_{sc}}$ and
$A_{V_{ce}}$ are calibrated using measurements of  $A_{V_{sp}}$ in
chosen reference fields, which means that $A_{V_{sc}}$ and $A_{V_{ce}}$ are
then not completely independent of $A_{V_{sp}}$. Note though, that the
calibration procedures only force methods to agree at a limited number of
points, and it does not force them to have the same structure or scale through
the cuts. $A_{V_{ISSA}}$ depends on a conversion from dust opacity at 100
\micron \/ to visual extinction, which ultimately relies on star count data in
Jarrett et al. (1989). Thus, $A_{V_{ISSA}}$ is {\it independent} of
any of the other extinction methods in this study, but it
is tied to the star count data of Jarrett et al. (1989).

Table 4 outlines the advantages and disadvantages, and the random and
systematic errors, of each of the four different methods of measuring
extinction used in this Paper.  

In principle, it would seem that the best method
to calculate extinction is using the color excess of individual stars with
measured spectral type, but this method is not without problem. One
inconvenience is the fact that
$R_V$ could have different values for different lines of sight. But systematic
errors due to unknown constants are also present in the other methods
($\beta$ for $A_{V_{ISSA}}$, $A_V/A_R$ for $A_{V_{sc}}$, and
$A_V/E_{V-R}$ for $A_{V_{ce}}$). Therefore not knowing the specific value of
$R_V$ for every line
of sight observed is not a disadvantage over
the other methods. The real drawback of this technique is the large amount
of time required to measure each and every star's spectrum. Thus,
although using the color excess of background stars with known spectral type
is the most direct and accurate way of measuring the extinction,
it is a very time consuming procedure
and it measures the extinction in a spatially nonuniform
fashion.

To asses the robustness of the four methods of measuring extinction used in
this Paper, we constructed plots of
$A_{V_{ISSA}}$, $A_{V_{sc}}$, and $A_{V_{ce}}$ versus $A_{V_{sp}}$,
at each point where all four methods can be used.
We do this in order to obtained a least square
fit for each of the three remaining methods plotted against $A_{V_{sp}}$.
The four points with the highest extinction  were not included in the fits,
since these are points near the extinction peaks of dark clouds.
It is clear that for these four points $A_{V_{sp}}$ is larger than any of
the other 3 methods since the 5\arcmin \/
beam of the other methods
does not resolve the extinction gradients in this regions of high extinction
(see section 4.1). When we constrain the fits to have a zero intercept, we find a
slope of one (within the errors) in all three comparisons. Thus, we believe all
four methods of obtaining extinction are robust.

\section{Summary and Conclusions}
We studied the extinction of a region  of Taurus in four different ways:
using the color excesses of background stars for which we had spectral types;
using the ISSA 60 and 100 \micron \/ images; using star counting;
and using an optical ($V$ and $R$) version of the average color
excess technique of (Lada et al. 1994).
All four give generally similar results. Therefore, any of the methods
discussed above can be used to obtain reliable information about the
extinction in regions where $A_V \lesssim 4$ mag.

We inter-compared the ISSA extinction and the extinction measured using
individual stars,
to study the spatial fluctuations in the dust distribution.
Excluding areas where there are extinction gradients
due to filamentary dark clouds
and IRAS cores,
we do not {\it detect} any variations in the structure on scales
smaller than 0.2 pc. With this result we are able to place a constraint on
the magnitude of the fluctuations. We conclude that in the regions
with $0.9 < A_V < 3$ mag, away from filamentary dark clouds
and IRAS cores,
there are no fluctuations
in the dust column density greater than
45\% (at the 99.7\% confidence level),
on scales smaller than 0.2 pc. On the other hand, in regions of
high extinction in the vicinity of dark clouds and IRAS cores, we do detect
statistically significant
deviations from the mean in dust column density on scales smaller than 0.2 pc.
Although it is not possible to definitively determine
 the nature of the fluctuations
with our data alone, the results of other studies
(Lada et al. 1998, Thoraval et al. 1997) and ours taken together
strongly favor unresolved steep gradients in clouds
over random fluctuations on the dust distribution.

A discrepancy between the extinction obtained through star
counting and average color excess
and the rest of the techniques
in the vicinity of R.A. 4h 19m, Dec. 27\arcdeg30\arcmin \/ (B1950),
leads us to believe in the existence of a previously unknown open stellar
cluster in the region.

\acknowledgements
We would like to thank Lucas M. Macri very much for his great help in the
photometry analysis and
taking time off his observing round to take a few frames for us in October
1996. We would also
like to thank Elizabeth Barton and Warren Brown for taking the November
1996 and March 1997 frames. A special thanks goes to Perry Berlind for
helping us with the acquisition of the spectral data. We would like to give
special thanks to our referee Dan Clemens, as well as to Douglas Finkbeiner
for their
very helpful comments and thorough critique of the paper.
Also thanks to Scott J. Kenyon,
George Field, Charlie Lada, and John Huchra for their
helpful remarks and Ian Reid for sharing his star
count Galaxy model with us.

\clearpage

\begin{deluxetable}{ccccccccccccc}
\scriptsize
\tablecolumns{13}
\tablecaption{Photometric and Spectroscopic Data
\label{photab}}
\tablehead{
\colhead{Star} & \colhead{R.A.} & \colhead{Dec} & \colhead{V} &
\colhead{$\sigma(V)$}
& \colhead{\bv} & \colhead{$\sigma(\bv)$} &
\colhead{Spectral} & \colhead{$E_{B-V}$} & \colhead{$A_{V}$}
& \colhead{$\sigma(A_{V})$} & \colhead{distance} & \colhead{$\sigma(d)$}
\\
\colhead{Name} & \colhead{(J2000)} & \colhead{(J2000)} & \colhead{[mag]} &
\colhead{[mag]} & \colhead{[mag]} & \colhead{[mag]} & \colhead{Type} &
\colhead{[mag]} &
\colhead{[mag]} & \colhead{[mag]} & \colhead{[pc]} & \colhead{[pc]}
}
\startdata

011005 & 4h22m39.3s & 22\arcdeg28\arcmin08\arcsec & 14.94 & 0.02 & 0.58 &
0.03 & F2 & 0.23 & 0.73 & 0.16 & 1669 & 261\nl
011003 & 4 22 38.5 & 22 38 47 & 16.57 & 0.02 & 0.97 & 0.04 & G8 & 0.24 &
0.75 & 0.22 & 1156 & 291\nl
011001 & 4 22 35.5 & 22 45 39 & 14.87 & 0.02 & 0.62 & 0.03 & F6 & 0.16 &
0.51 & 0.16 & 1295 & 202\nl
021015 & 4 22 35.5 & 22 48 02 & 15.99 & 0.02 & 0.83 & 0.03 & F8 & 0.31 &
0.97 & 0.18 & 1459 & 235\nl
021014 & 4 22 36.2 & 22 53 01 & 15.81 & 0.02 & 1.07 & 0.03 & G8 & 0.34 &
1.05 & 0.21 &  711 & 177\nl
021013 & 4 22 36.9 & 22 57 46 & 15.41 & 0.02 & 0.69 & 0.06 & F6 & 0.23 &
0.72 & 0.23 & 1511 & 262\nl
021012 & 4 22 39.5 & 23 03 30 & 14.94 & 0.02 & 0.71 & 0.03 & F7 & 0.21 &
0.64 & 0.16 &  955 & 149\nl
021011 & 4 22 36.3 & 23 05 39 & 15.18 & 0.02 & 0.74 & 0.03 & F7 & 0.24 &
0.73 & 0.16 & 1231 & 192\nl
00N5.2 & 4 22 34.4 & 23 08 09 & 14.40 & 0.02 & 0.58 & 0.02 & F3 & 0.21 &
0.65 & 0.15 & 1292 & 199\nl
031025 & 4 22 33.4 & 23 10 38 & 15.37 & 0.02 & 0.84 & 0.03 & G0 & 0.26 &
0.81 & 0.18 & 1075 & 263\nl
031024 & 4 22 29.7 & 23 12 54 & 15.47 & 0.02 & 0.90 & 0.03 & G4 & 0.26 &
0.79 & 0.19 &  861 & 212\nl
031023 & 4 22 29.7 & 23 16 11 & 15.23 & 0.02 & 0.69 & 0.03 & F6 & 0.23 &
0.70 & 0.16 & 1398 & 219\nl
031022 & 4 22 29.9 & 23 19 32 & 16.34 & 0.02 & 0.94 & 0.03 & F9 & 0.39 &
1.21 & 0.18 & 1462 & 236\nl
TDC071 & 4 22 22.5 & 23 20 07 & 15.60 & 0.02 & 0.54 & 0.03 & A9 & 0.27 &
0.82 & 0.17 & 2720 & 434\nl
041035 & 4 22 33.7 & 23 28 24 & 14.57 & 0.04 & 0.78 & 0.07 & F7 & 0.28 &
0.86 & 0.25 &  877 & 158\nl
TDC085 & 4 22 18.6 & 23 33 10 & 16.25 & 0.05 & 0.81 & 0.07 & F6 & 0.34 &
1.07 & 0.25 & 1884 & 342\nl
041033 & 4 22 33.6 & 23 37 38 & 16.23 & 0.04 & 1.17 & 0.06 & G5 & 0.51 &
1.59 & 0.27 &  809 & 212\nl
041032 & 4 22 32.4 & 23 41 30 & 15.01 & 0.04 & 0.98 & 0.06 & F7 & 0.48 &
1.50 & 0.23 &  798 & 139\nl
041031 & 4 22 35.3 & 23 45 31 & 16.16 & 0.04 & 1.50 & 0.06 & F8 & 0.99 &
3.05 & 0.25 &  605 & 109\nl
051045 & 4 22 30.2 & 23 47 47 & 15.15 & 0.04 & 1.07 & 0.06 & F5 & 0.65 &
2.02 & 0.23 &  805 & 140\nl
051044 & 4 22 34.6 & 23 52 36 & 14.90 & 0.04 & 0.81 & 0.06 & F3 & 0.44 &
1.38 & 0.23 & 1163 & 202\nl
TDC101 & 4 22 44.6 & 23 53 49 & 15.51 & 0.04 & 0.75 & 0.06 & F5 & 0.33 &
1.02 & 0.23 & 1508 & 263\nl
051043 & 4 22 37.8 & 23 55 57 & 16.08 & 0.02 & 0.77 & 0.03 & F6 & 0.31 &
0.96 & 0.16 & 1839 & 288\nl
051042 & 4 22 36.0 & 23 59 45 & 15.02 & 0.02 & 0.79 & 0.02 & F6 & 0.33 &
1.01 & 0.15 & 1103 & 170\nl
051041 & 4 22 36.8 & 24 03 60 & 14.37 & 0.02 & 0.90 & 0.02 & F9 & 0.35 &
1.08 & 0.17 &  626 &  99\nl
061055 & 4 22 38.7 & 24 12 10 & 16.43 & 0.04 & 0.84 & 0.06 & F6 & 0.38 &
1.19 & 0.22 & 1943 & 334\nl
061054 & 4 22 43.4 & 24 15 44 & 15.29 & 0.04 & 0.87 & 0.05 & F7 & 0.37 &
1.14 & 0.21 & 1072 & 181\nl
061053 & 4 22 39.6 & 24 19 04 & 14.94 & 0.04 & 0.74 & 0.05 & F0 & 0.43 &
1.32 & 0.21 & 1530 & 259\nl
061052 & 4 22 35.9 & 24 21 58 & 15.83 & 0.04 & 1.03 & 0.06 & G1 & 0.43 &
1.34 & 0.24 &  951 & 243\nl
061051 & 4 22 37.1 & 24 24 34 & 15.09 & 0.04 & 1.42 & 0.05 & G9 & 0.64 &
1.98 & 0.26 &  302 &  78\nl
071065 & 4 22 39.4 & 24 29 30 & 16.04 & 0.04 & 1.02 & 0.06 & F7 & 0.52 &
1.63 & 0.22 & 1207 & 208\nl
071064 & 4 22 37.6 & 24 34 02 & 12.93 & 0.02 & 0.86 & 0.02 & F5 & 0.44 &
1.37 & 0.15 &  390 &  60\nl
071063 & 4 22 32.9 & 24 37 16 & 14.63 & 0.04 & 0.95 & 0.05 & F9 & 0.40 &
1.25 & 0.23 &  655 & 114\nl
N14.51 & 4 22 41.9 & 24 37 49 & 14.15 & 0.02 & 0.77 & 0.02 & A9 & 0.50 &
1.57 & 0.16 &  991 & 155\nl
N14.52 & 4 22 58.2 & 24 40 35 & 15.25 & 0.02 & 0.78 & 0.03 & F6 & 0.32 &
1.00 & 0.16 & 1231 & 192\nl
071062 & 4 22 40.3 & 24 41 14 & 12.82 & 0.02 & 1.34 & 0.02 & G9 & 0.56 &
1.74 & 0.30 &  199 &  32\nl
071061 & 4 22 40.3 & 24 45 56 & 15.14 & 0.04 & 0.85 & 0.05 & F3 & 0.48 &
1.49 & 0.21 & 1231 & 208\nl
081075 & 4 22 34.3 & 24 49 39 & 15.40 & 0.04 & 1.31 & 0.05 & G9 & 0.53 &
1.65 & 0.26 &  406 & 105\nl
081074 & 4 22 31.2 & 24 53 44 & 15.78 & 0.04 & 0.93 & 0.05 & F0 & 0.62 &
1.93 & 0.21 & 1703 & 288\nl
AG-136 & 4 22 44.0 & 24 57 04 & 12.32 & 0.02 & 0.93 & 0.02 & F7 & 0.43 &
1.33 & 0.15 &  249 &  38\nl
081073 & 4 22 35.2 & 24 57 59 & 15.66 & 0.04 & 1.12 & 0.05 & F7 & 0.62 &
1.93 & 0.21 &  883 & 149\nl
AG-133 & 4 21 48.0 & 25 05 20 & 14.15 & 0.03 & 0.91 & 0.04 & F7 & 0.41 &
1.27 & 0.18 &  596 &  96\nl
AG-132 & 4 21 31.0 & 25 05 48 & 14.17 & 0.02 & 1.14 & 0.02 & G1 & 0.54 &
1.67 & 0.19 &  379 &  93\nl
091085 & 4 22 30.3 & 25 09 19 & 16.96 & 0.04 & 2.12 & 0.09 & G0 & 1.54 &
4.78 & 0.33 &  358 &  99\nl
AG-102 & 4 22 49.0 & 25 11 23 & 12.29 & 0.03 & 1.27 & 0.10 & A9 & 1.00 &
3.10 & 0.34 &  207 &  43\nl
091084 & 4 22 31.0 & 25 12 40 & 15.91 & 0.04 & 1.36 & 0.06 & G3 & 0.73 &
2.27 & 0.24 &  586 & 150\nl
091083 & 4 22 36.5 & 25 15 57 & 16.32 & 0.04 & 1.23 & 0.06 & G9 & 0.45 &
1.41 & 0.27 &  696 & 182\nl
092087 & 4 21 26.0 & 25 19 49 & 15.75 & 0.02 & 0.85 & 0.03 & A2 & 0.79 &
2.45 & 0.17 & 2158 & 343\nl
00Ld.1 & 4 21 31.5 & 25 20 39 & 15.41 & 0.02 & 0.53 & 0.03 & A2 & 0.47 &
1.46 & 0.17 & 2949 & 469\nl
AG-105 & 4 21 28.0 & 25 20 51 & 15.16 & 0.03 & 1.29 & 0.04 & G9 & 0.51 &
1.58 & 0.24 &  376 &  96\nl
091081 & 4 22 34.2 & 25 25 32 & 14.22 & 0.02 & 0.78 & 0.02 & F5 & 0.36 &
1.11 & 0.15 &  795 & 123\nl
101095 & 4 22 32.0 & 25 27 51 & 15.27 & 0.02 & 0.85 & 0.03 & F5 & 0.43 &
1.34 & 0.16 & 1162 & 182\nl
0N20.1 & 4 22 43.0 & 25 28 18 & 16.14 & 0.03 & 0.78 & 0.05 & F6 & 0.32 &
1.00 & 0.19 & 1854 & 304\nl
0N20.2 & 4 22 58.9 & 25 28 32 & 16.34 & 0.03 & 0.40 & 0.05 & A2 & 0.34 &
1.05 & 0.20 & 5460 & 911\nl
101094 & 4 22 30.5 & 25 32 13 & 13.61 & 0.02 & 0.89 & 0.02 & F6 & 0.43 &
1.35 & 0.15 &  493 &  76\nl
101092 & 4 22 24.3 & 25 42 15 & 15.55 & 0.02 & 1.01 & 0.04 & F7 & 0.51 &
1.57 & 0.17 &  986 & 157\nl
101091 & 4 22 30.3 & 25 44 58 & 13.27 & 0.02 & 0.74 & 0.02 & A8 & 0.50 &
1.57 & 0.16 &  692 & 108\nl
0N22.1 & 4 22 28.9 & 25 46 24 & 14.61 & 0.02 & 0.64 & 0.02 & A5 & 0.50 &
1.55 & 0.16 & 1553 & 244\nl
111105 & 4 22 32.8 & 25 48 49 & 16.85 & 0.02 & 0.64 & 0.04 & A6 & 0.48 &
1.49 & 0.18 & 4279 & 691\nl
111104 & 4 22 34.0 & 25 51 39 & 14.83 & 0.02 & 0.88 & 0.02 & F5 & 0.46 &
1.41 & 0.15 &  918 & 142\nl
111102 & 4 22 30.2 & 25 58 31 & 15.90 & 0.02 & 1.24 & 0.03 & G9 & 0.46 &
1.43 & 0.22 &  567 & 142\nl
111101 & 4 22 38.2 & 26 02 53 & 12.76 & 0.05 & 1.18 & 0.10 & G1 & 0.58 &
1.80 & 0.35 &  186 &  52\nl
121115 & 4 22 37.4 & 26 08 34 & 14.27 & 0.04 & 1.17 & 0.06 & G1 & 0.57 &
1.78 & 0.22 &  378 &  97\nl
121113 & 4 22 36.1 & 26 15 12 & 15.64 & 0.04 & 0.93 & 0.06 & F6 & 0.47 &
1.45 & 0.22 & 1197 & 206\nl
121112 & 4 22 37.5 & 26 21 20 & 16.03 & 0.04 & 1.18 & 0.06 & G8 & 0.45 &
1.41 & 0.27 &  665 & 174\nl
131124 & 4 22 45.2 & 26 33 04 & 15.09 & 0.04 & 1.12 & 0.06 & G0 & 0.54 &
1.66 & 0.24 &  639 & 163\nl
131123 & 4 22 42.9 & 26 35 57 & 15.37 & 0.04 & 1.23 & 0.06 & G3 & 0.60 &
1.85 & 0.26 &  554 & 144\nl
131122 & 4 22 45.2 & 26 40 32 & 16.44 & 0.04 & 1.29 & 0.07 & F7 & 0.79 &
2.44 & 0.26 & 1002 & 183\nl
131121 & 4 22 41.8 & 26 42 28 & 15.61 & 0.04 & 0.96 & 0.06 & F3 & 0.59 &
1.83 & 0.23 & 1307 & 228\nl
141135 & 4 22 39.7 & 26 50 00 & 15.79 & 0.02 & 1.24 & 0.03 & F7 & 0.74 &
2.29 & 0.16 &  795 & 124\nl
141134 & 4 22 33.9 & 26 51 38 & 14.29 & 0.02 & 1.34 & 0.02 & A5 & 1.20 &
3.73 & 0.16 &  492 &  77\nl
141133 & 4 22 38.1 & 26 55 37 & 15.08 & 0.02 & 0.94 & 0.03 & F3 & 0.57 &
1.76 & 0.15 & 1056 & 163\nl
141131 & 4 22 39.8 & 27 04 44 & 16.05 & 0.02 & 0.86 & 0.04 & F4 & 0.47 &
1.47 & 0.17 & 1722 & 273\nl
151145 & 4 22 33.1 & 27 09 43 & 17.32 & 0.03 & 1.06 & 0.05 & F7 & 0.56 &
1.73 & 0.19 & 2074 & 341\nl
00B2.2 & 4 21 25.5 & 27 11 31 & 13.89 & 0.02 & 0.72 & 0.02 & A2 & 0.66 &
2.03 & 0.16 & 1125 & 176\nl
00B1.2 & 4 21  0.7 & 27 11 16 & 16.01 & 0.03 & 0.58 & 0.04 & F1 & 0.25 &
0.79 & 0.18 & 2916 & 470\nl
TDC301 & 4 22 18.4 & 27 12 20 & 17.35 & 0.03 & 1.16 & 0.05 & G4 & 0.52 &
1.60 & 0.25 & 1412 & 364\nl
TDC304 & 4 22 34.0 & 27 13 55 & 15.50 & 0.02 & 1.13 & 0.03 & F9 & 0.58 &
1.79 & 0.19 &  761 & 124\nl
00B2.1 & 4 21 22.0 & 27 16 28 & 13.01 & 0.02 & 0.47 & 0.02 & A1 & 0.44 &
1.35 & 0.16 & 1179 & 185\nl
TDC312 & 4 22 11.5 & 27 21 25 & 15.72 & 0.02 & 0.92 & 0.03 & F6 & 0.46 &
1.44 & 0.16 & 1247 & 195\nl
151141 & 4 22 30.4 & 27 24 23 & 16.62 & 0.02 & 1.40 & 0.04 & G4 & 0.76 &
2.37 & 0.22 &  708 & 178\nl
TDC321 & 4 22 28.2 & 27 25 22 & 15.29 & 0.02 & 1.20 & 0.03 & A9 & 0.93 &
2.87 & 0.17 &  919 & 146\nl
TDC323 & 4 22 15.2 & 27 26 41 & 16.67 & 0.02 & 1.03 & 0.04 & G3 & 0.40 &
1.23 & 0.20 & 1342 & 333\nl
161152 & 4 22 43.4 & 27 42 02 & 14.42 & 0.02 & 1.03 & 0.03 & A8 & 0.80 &
2.47 & 0.17 &  776 & 123\nl
TDC332 & 4 22 34.2 & 27 42 45 & 16.15 & 0.02 & 1.17 & 0.03 & F5 & 0.75 &
2.32 & 0.16 & 1110 & 174\nl
171165 & 4 22 33.0 & 27 48 09 & 14.61 & 0.02 & 1.36 & 0.03 & G5 & 0.70 &
2.17 & 0.21 &  294 &  73\nl
TDC342 & 4 22 38.7 & 27 49 55 & 15.35 & 0.02 & 0.92 & 0.03 & F2 & 0.57 &
1.76 & 0.16 & 1250 & 196\nl
171164 & 4 22 31.9 & 27 51 36 & 15.52 & 0.02 & 1.41 & 0.03 & G5 & 0.75 &
2.32 & 0.21 &  417 & 104\nl
TDC341 & 4 22 47.1 & 27 52 42 & 15.78 & 0.02 & 0.83 & 0.03 & F0 & 0.52 &
1.60 & 0.16 & 1974 & 310\nl
TDC351 & 4 22 15.5 & 28 00 46 & 15.48 & 0.02 & 0.82 & 0.03 & F4 & 0.43 &
1.33 & 0.16 & 1415 & 221\nl
TDC354 & 4 22 42.8 & 28 00 48 & 15.89 & 0.02 & 0.99 & 0.03 & F6 & 0.53 &
1.64 & 0.16 & 1230 & 192\nl
TDC361 & 4 22 16.9 & 28 04 22 & 15.12 & 0.02 & 0.80 & 0.03 & F1 & 0.47 &
1.44 & 0.16 & 1431 & 224\nl
TDC364 & 4 22 50.4 & 28 12 00 & 16.25 & 0.02 & 1.15 & 0.03 & G6 & 0.47 &
1.46 & 0.22 &  825 & 208\nl
TDC373 & 4 22 25.0 & 28 17 25 & 15.07 & 0.02 & 0.81 & 0.03 & F4 & 0.42 &
1.29 & 0.16 & 1192 & 186\nl
181172 & 4 22 35.2 & 28 18 55 & 15.37 & 0.02 & 0.80 & 0.03 & F7 & 0.30 &
0.94 & 0.16 & 1222 & 191\nl

\enddata

\end{deluxetable}

\clearpage

\begin{deluxetable}{crrrr}
\tablecolumns{5}
\tablecaption{Background Subtraction for ISSA images
\label{irastab}}
\tablehead{
\colhead{} &
\multicolumn{2}{c}{Minimum Flux} &
\multicolumn{2}{c}{Background Subtraction Constant}
\\  \cline{2-3}   \cline{4-5}
\colhead{} & \colhead{100 \micron} & \colhead{60 \micron}
& \colhead{100 \micron \/ Flux} & \colhead{60 \micron \/ Flux}
\\
\colhead{Image} & \colhead{(MJy Sr$^{-1}$)} & \colhead{(MJy Sr$^{-1}$)} &
\colhead{(MJy Sr$^{-1}$)} & \colhead{(MJy Sr$^{-1}$)}
}
\startdata
north\tablenotemark{a} & 8.89 & $-0.57$ & 8.50 & $-0.70$\nl
south\tablenotemark{b} & 7.27 & $-1.05$ & 6.60 & $-1.04$\enddata
\tablenotetext{a}{The north images come from the ISSA images i311b3h0 and
i311b4h0, which are centered at 4h36m, 30\arcdeg \/ (B1950)}
\tablenotetext{b}{The south images come from the ISSA images i276b3h0 and
i276b4h0, which are centered at
4h12m, 20\arcdeg \/ (B1950)}

\end{deluxetable}

\clearpage

\begin{deluxetable}{ccrr}
\tablecolumns{4}
\tablecaption{Control fields for star count
\label{starctab}}
\tablehead{
\colhead{Star\tablenotemark{a}} & \colhead{$N_{ref}$} & \colhead{$A_{V_{ref}}$
[mag]} & \colhead{Range\tablenotemark{b}}
}
\startdata
011001 & 41 & $0.50 \pm 0.2$ & $-18.9\arcdeg\leq b^{II}< -17.9\arcdeg$\nl
051043 & 36 & $1.00 \pm 0.2$ & $-17.9\arcdeg\leq b^{II}< -16.9\arcdeg$\nl
101095 & 34 & $1.30 \pm 0.2$ & $-16.9\arcdeg\leq b^{II}< -15.9\arcdeg$\nl
151145 & 28 & $1.70 \pm 0.2$ & $-15.9\arcdeg\leq b^{II}< -14.9\arcdeg$\enddata
\tablenotetext{a}{Each reference field is tied to only one spectroscopic star}
\tablenotetext{b}{The star count extinction in the given range of $b^{II}$
is calibrated using the reference field of the respective star.}

\end{deluxetable}

\newpage
\clearpage
\thispagestyle{empty}

\begin{figure}
\vspace{6.5in}
\centerline{
\includegraphics{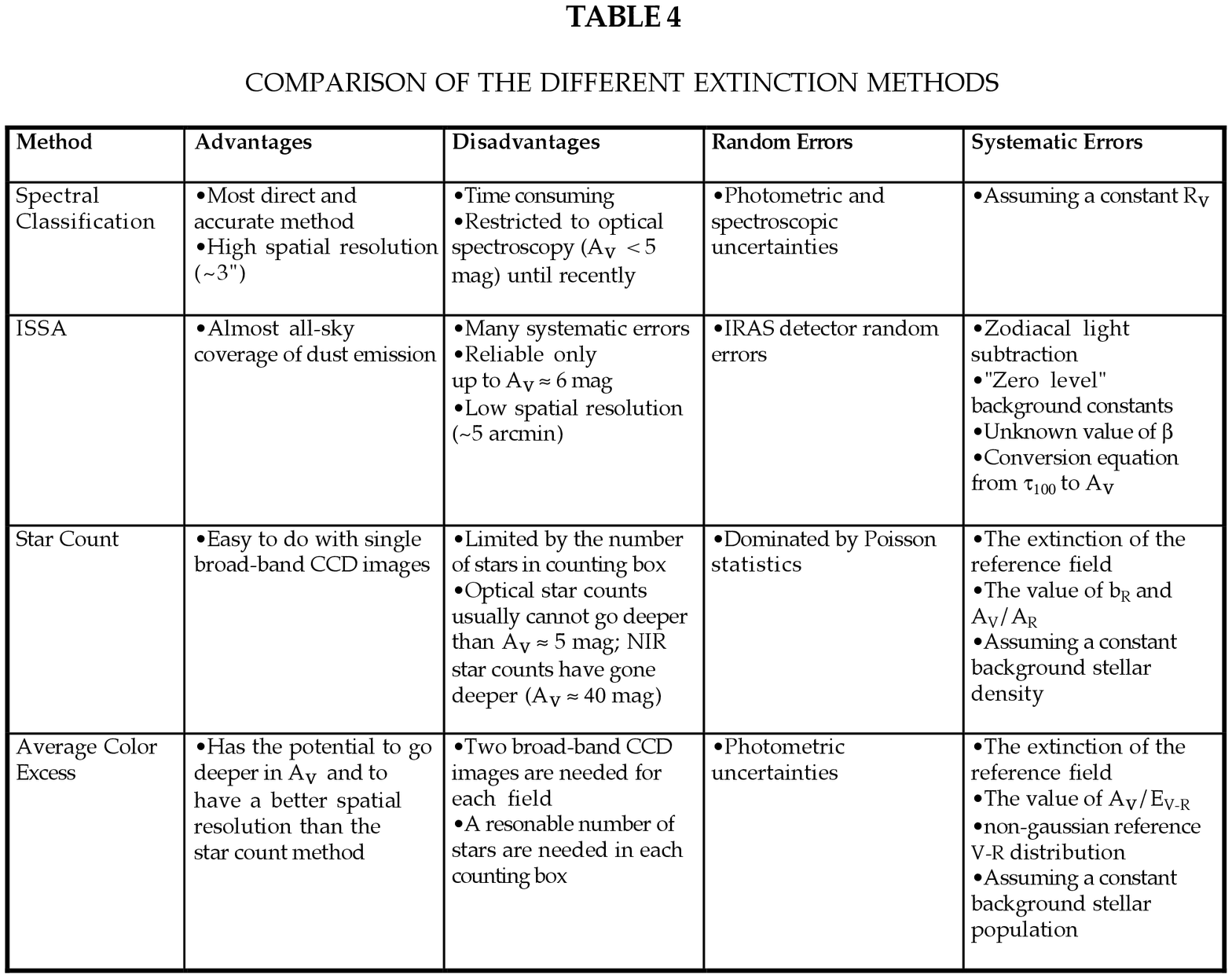}
}
\end{figure}

\newpage
\clearpage
\thispagestyle{empty}

\begin{figure}
\vspace{6.5in}
\centerline{
\includegraphics{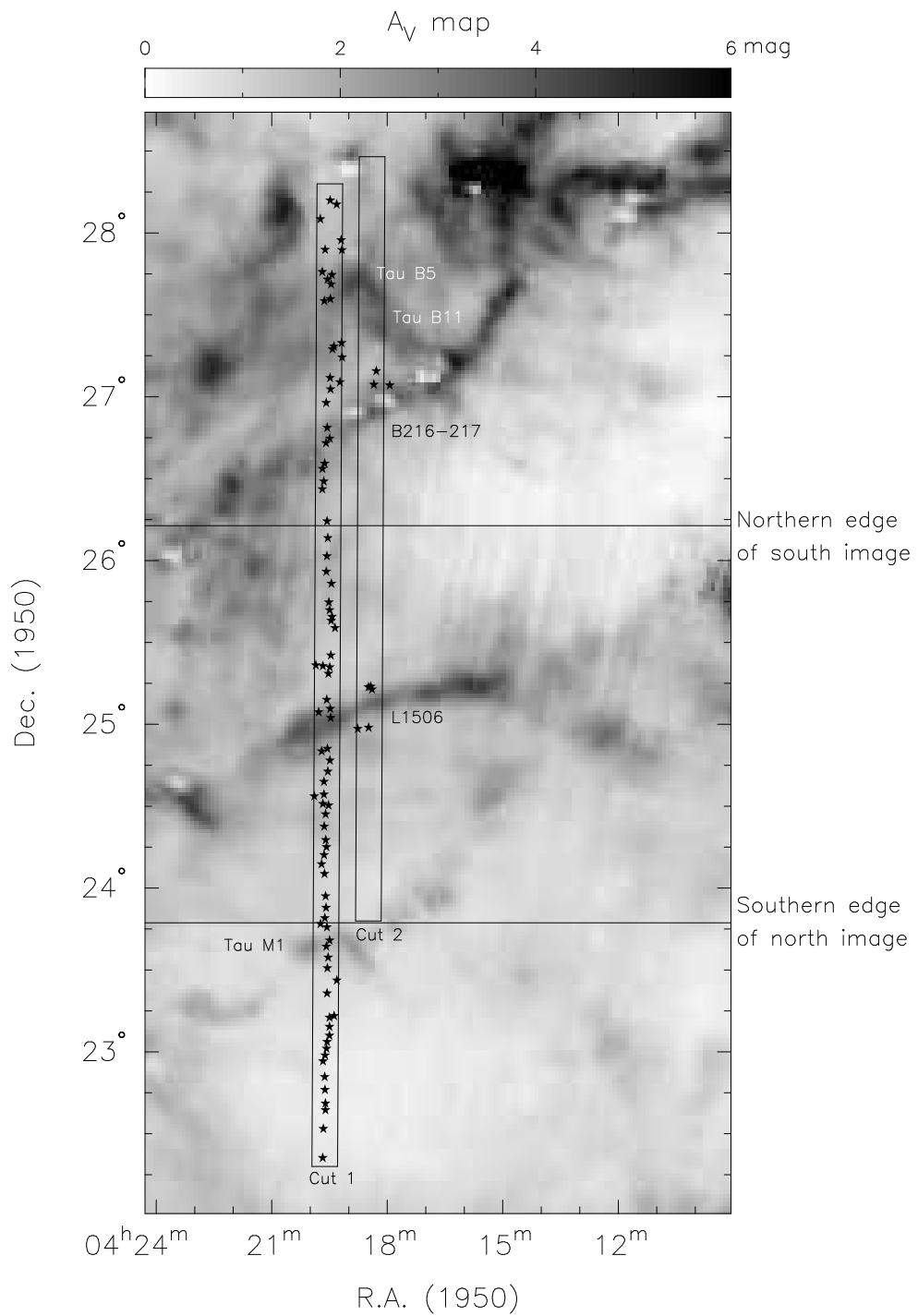}
}
\caption{Extinction ($A_V$) map of part of the Taurus
dark cloud region. The map was obtained using the method described in \S 2.3.
The two cuts for which we have photometric (CCD) data are shown.
The two filamentary dark clouds L1506 and B216-217 and the
three  IRAS cores Tau M1, Tau B5 and Tau B11 are identified.
The star symbols represent the position of the stars that were classified
by spectra.
The horizontal lines mark the northern and southern edges of the two images
that were used to make this map.}

\end{figure}

\newpage
\clearpage
\thispagestyle{empty}

\begin{figure}
\vspace{6.5in}
\centerline{
\includegraphics{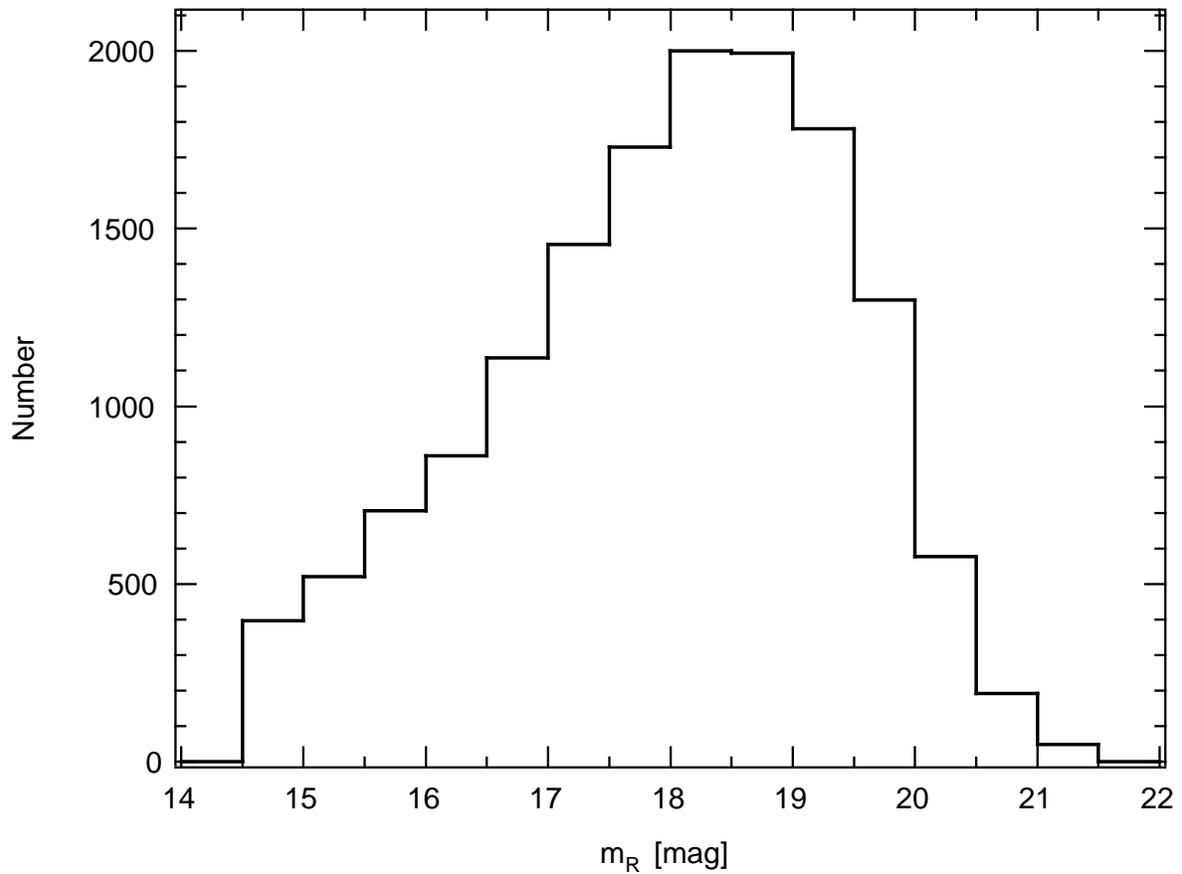}
}
\caption{A histogram of the distribution of stars with respect
to apparent $R$-band magnitude ($m_R$) is shown. These are the stars
we detected in the 64 $R$-band frames we took in November 1995. 
Stars with $m_R < 14.5$ are saturated
in the 200 sec exposures, and thus we do not include them in the distribution. 
From this figure we estimate our sample is at least 90\%  
complete for $14.5 \leq m_R < 18.0$.}

\end{figure}
\newpage
\clearpage
\thispagestyle{empty}
\begin{figure}
\vspace{5.5in}
\centerline{
\includegraphics{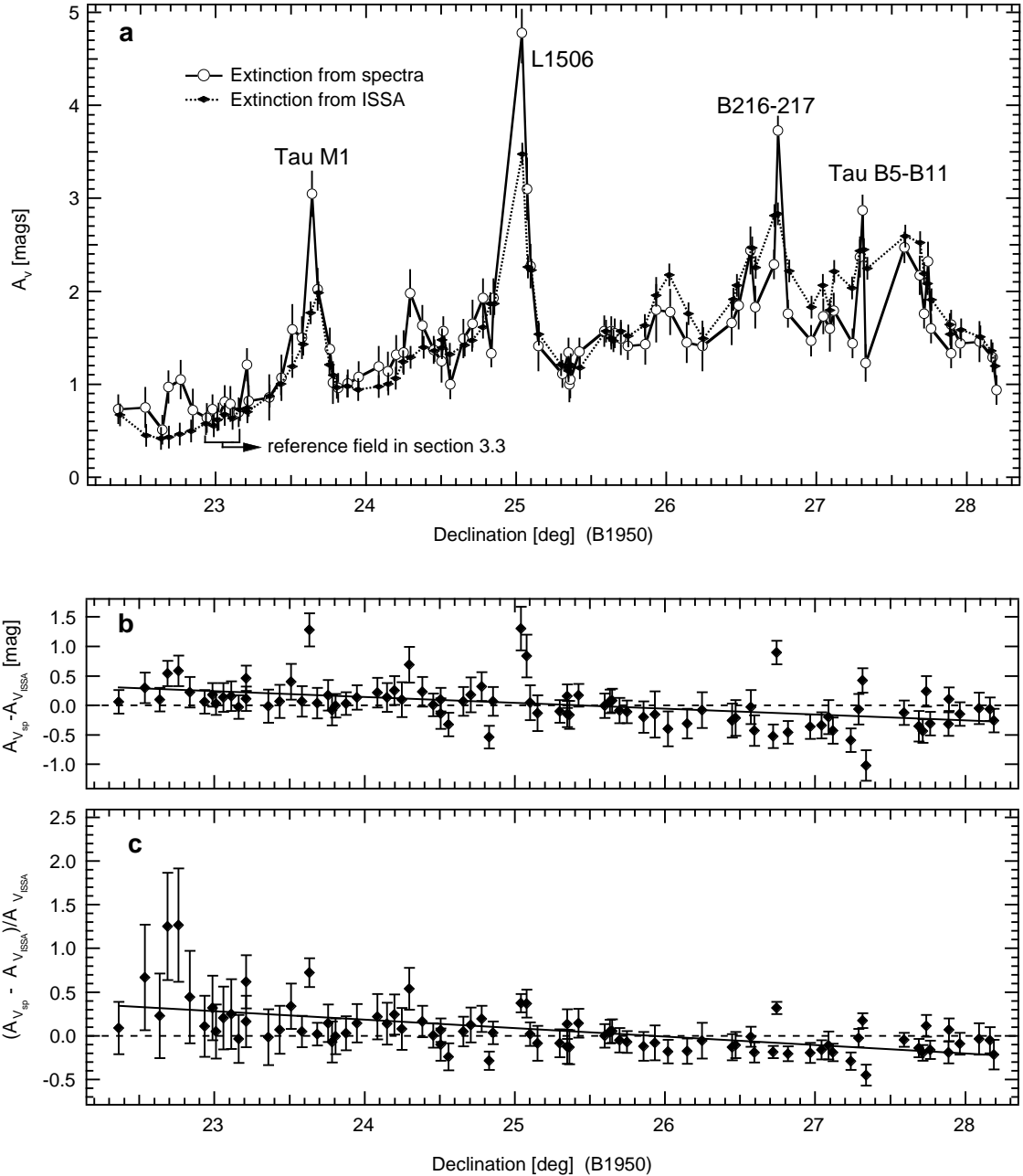}
}
\caption{The top panel (Figure 3a) shows the extinction of every star we
classified by spectral type with distances larger then 140 pc in cut 1,
as well as  the extinction at the same position on the sky using the
ISSA extinction map (Figure 1). The error bars shown represent
only the random erors for each of the two methods, systematic errors
are not included.
Also shown in Figure 3a is the region used as reference
field for the extinction obtained using the technique described
in section 3.3.
Figure 3b shows the difference between
the two different extinction values from Figure 3a. Figure 3c shows
the difference (from Figure 3b) divided by the ISSA extinction ($A_{V_{ISSA}}$).
The solid lines in Figures 3b and 3c are the unweighted linear
fits to the points in each of the two plots. These lines indicate the small
gradient in extinction with declination in $A_{V_{ISSA}}$, which we believe
is due to an imperfect zodiacal light subtraction of the ISSA images.}

\end{figure}

\newpage
\clearpage
\thispagestyle{empty}

\begin{figure}
\vspace{6.0in}
\centerline{
\includegraphics{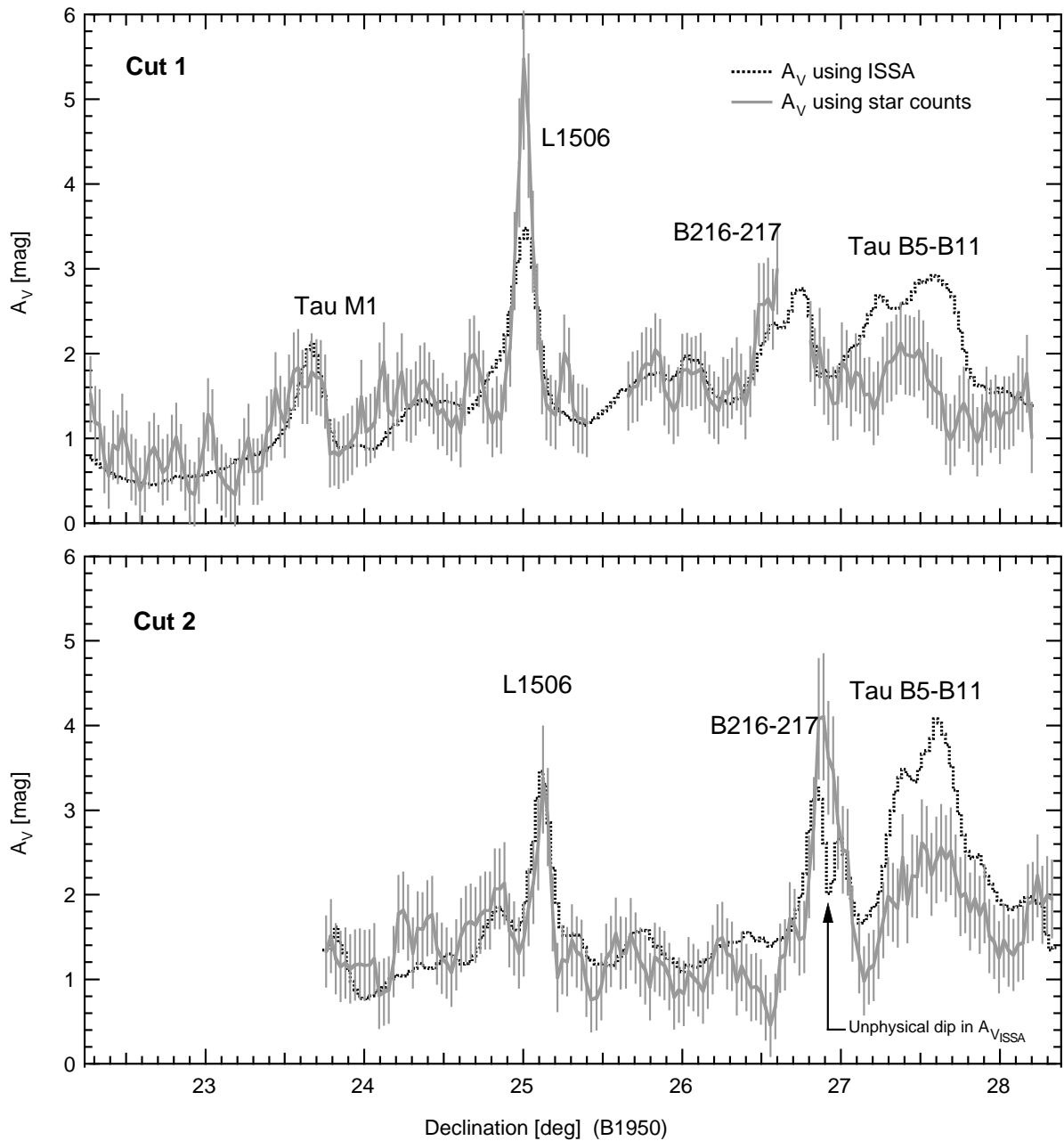}
}
\caption{Plots of extinction vs. declination for both
cut 1 and cut 2. The solid line is the extinction obtained through
star counts. The dotted line
is the extinction obtained using the ISSA 60 and 100 \micron \/ images.
Both methods are averaged over the 10\arcmin \/ width of the cut.
Breaks in the star count extinction trace are due to missing $R$-band
frames. The error bars in the star count extinction represent
the 1 $\sigma$ error.
The rise in extinction  associated with the dark clouds and
IRAS cores are identified. The dip in the extinction peak associated with
B216-217, in the $A_{V_{ISSA}}$ trace of cut 2 is due to an unreal drop in
the extinction caused by the presence of IRAS point source in the images.}

\end{figure}
\newpage
\clearpage
\thispagestyle{empty}
\begin{figure}
\vspace{6.5in}
\centerline{
\includegraphics{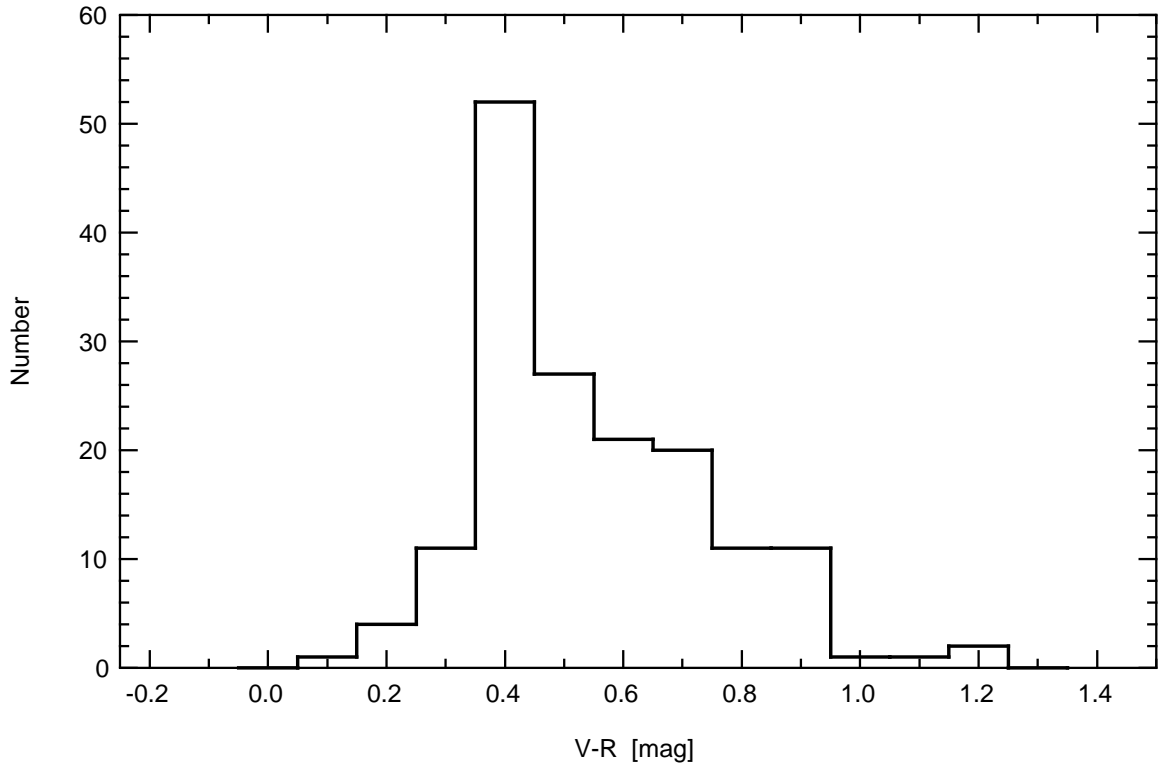}
}
\caption{Distribution of \vr \/ colors in the reference field
for the average color excess method described in \S 3.3.}

\end{figure}

\newpage
\clearpage
\thispagestyle{empty}

\begin{figure}
\vspace{6.0in}
\centerline{
\includegraphics{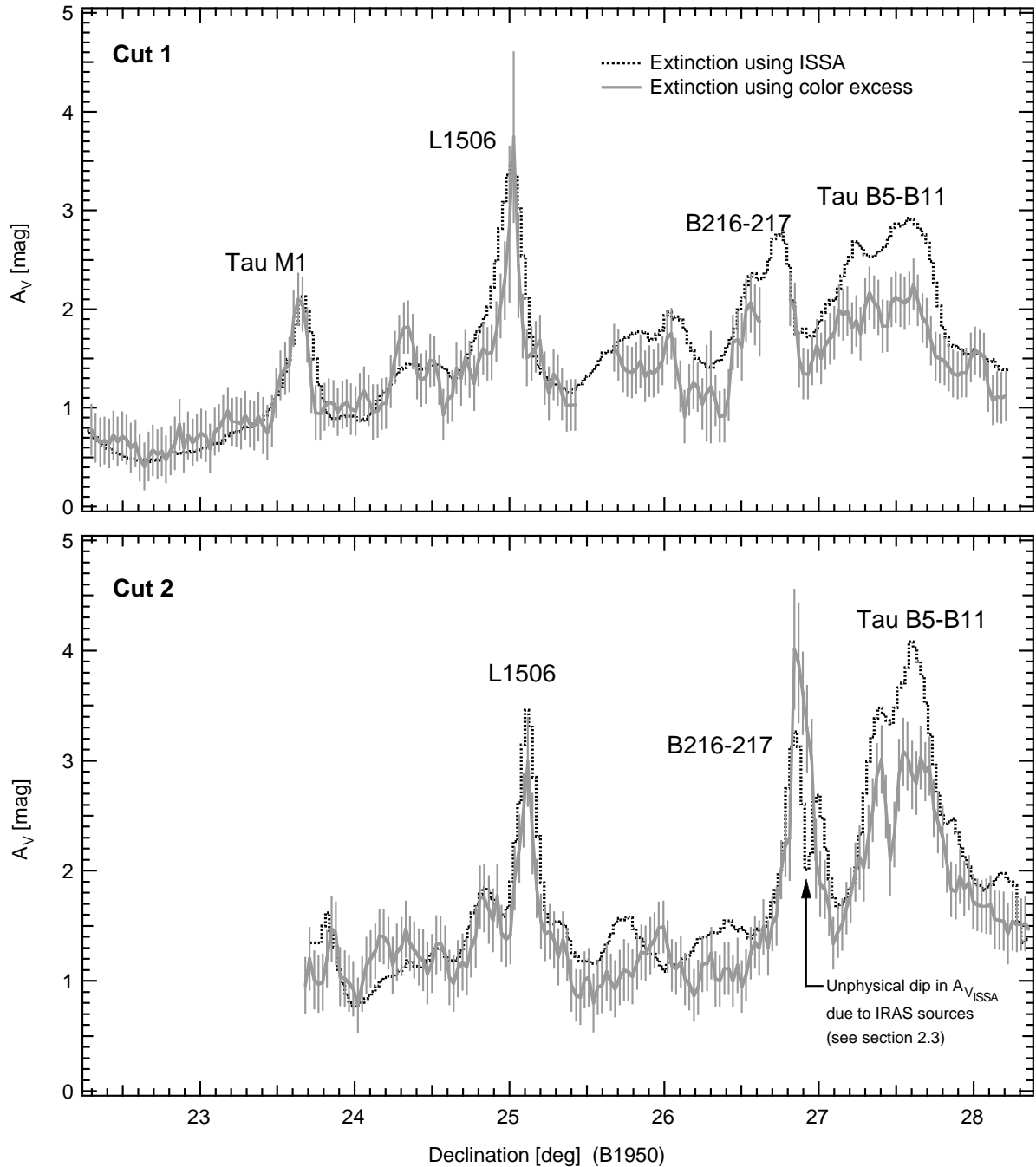}
}
\caption{Similar to Figure 4, but this time the solid line
is the extinction obtained through the average color excess method described
in \S 3.3. Breaks in the
solid line trace in cut 1 are due to missing $R$-band frames.
The error bars in the average color excess extinction represent
the 1 $\sigma$ error.
The dip in the extinction peak associated with
B216-217, in the $A_{V_{ISSA}}$ trace of cut 2 is due to an unreal drop in
the extinction caused by the presence of IRAS point source in the images.}

\end{figure}

\end{document}